\documentclass{naturemod}
\usepackage{graphicx}
\usepackage{graphics}
\usepackage{amssymb}
\usepackage{xcolor}
\usepackage{amsmath}
\usepackage{newtxtext}
\usepackage[varg,smallerops,upint]{newtxmath}
\usepackage{newfloat}
\usepackage{threeparttablex, tablefootnote}
\usepackage{longtable}
\usepackage{booktabs}
\usepackage{soul}
\usepackage{makecell} 
\usepackage{lineno}

\DeclareFloatingEnvironment[fileext=lop]{Extended Data Figure}

\newcommand{\mstar}{M_{\star}}
\newcommand{\msun}{M_{\odot}}
\newcommand{\mbh}{M_{\rm BH}}

\definecolor{ForestGreen}{RGB}{34,139,34}
\def\zy#1 {{\textcolor{ForestGreen}{#1}}\ }
\def\blue#1 {{\textcolor{blue}{#1}}\ }

\title{Overmassive Black holes live in compact galaxies in the early Universe}

\author{Yuxuan Wu$^{1,2}$, Tao Wang$^{1,2}$, Daizhong Liu$^{3}$, Qinghua Tan$^{3}$, Luis C. Ho$^{4,5}$, Zhiyu Zhang$^{1,2}$, Yong Shi$^{1,2}$, Ke Xu$^{1,2}$, Kotaro Kohno$^{6,7}$, Ran Wang$^{4,5}$, Takuma Izumi$^{8,9,10}$, Zhaozhou Li$^{11}$}

\begin{document}
\maketitle

\begin{spacing}{1.2}
\begin{affiliations}
\item School of Astronomy and Space Science, Nanjing University, Nanjing, Jiangsu 210093, China
\item Key Laboratory of Modern Astronomy and Astrophysics, Nanjing University, Ministry of Education, Nanjing 210093, China
\item Purple Mountain Observatory, Chinese Academy of Sciences, 10 Yuanhua Road, Nanjing 210023, China.
\item Kavli Institute for Astronomy and Astrophysics, Peking University, Beijing 100871, China
\item Department of Astronomy, School of Physics, Peking University, Beijing 100871, China
\item Institute of Astronomy, School of Science, The University of Tokyo, 2-21-1 Osawa, Mitaka, Tokyo 181-0015, Japan.
\item Research Center for the Early Universe, School of Science, The University of Tokyo, 7-3-1 Hongo, Bunkyo, Tokyo 113-0033, Japan.
\item National Astronomical Observatory of Japan, 2-21-1 Osawa, Mitaka, Tokyo 181-8588, Japan
\item Department of Astronomical Science, The Graduate University for Advanced Studies, SOKENDAI, 2-21-1 Osawa, Mitaka, Tokyo 181-8588, Japan 
\item Astronomical Science, The Graduate University for Advanced Studies, SOKENDAI, 2-21-1 Osawa, Mitaka, Tokyo 181-8588, Japan 
\item Centre for Astrophysics and Planetary Science, Racah Instituteof Physics, The Hebrew University, Jerusalem, 91904, Israel
\end{affiliations}
\begin{abstract}
A significant population of quasars have been found to exist within the first Gyr of cosmic time~\cite{Fan2006,Wang2013,Jiang2016,Decarli2018,Venemans2020}. Most of them have high black hole (BH) masses ($\mbh \sim 10^{8-10} \msun$) with an elevated BH-to-stellar mass ratio compared to typical local galaxies, posing challenges to our understanding of the formation of supermassive BHs and their coevolution with host galaxies~\cite{Reines2015,Kormendy2013}. Here, based on size measurements of [C {\sc ii}] 158$\mu$m emission for a statistical sample of $z \sim 6$ quasars, we find that their host galaxies are systematically more compact (with half-light radius $R_{\rm e} \sim 1.6$ kpc) than typical star-forming galaxies at the same redshifts. Specifically, the sizes of the most compact quasar hosts, which also tend to contain less cold gas than their more extended counterparts, are comparable to that of massive quiescent galaxies at $z \sim 4-5$. These findings reveal an intimate connection between the formation of massive BHs and compactness of their host galaxies in the early universe. These compact quasar hosts are promising progenitors of the first population of quiescent galaxies.

\end{abstract}


Since the first $z \sim 6$ quasars were discovered two decades ago, the formation mechanism of SMBHs with masses $\mbh \sim 10^{8-10} \msun$ in the early Universe has been an important question in extragalactic astronomy. To answer this question, it is essential to obtain detailed physical properties of the quasar host galaxies, which are crucial for understanding what kind of environment favors such rapid formation of SMBHs. However, even with JWST imaging, only the host galaxies of relatively faint quasars at these high redshifts could be marginally resolved in the ultraviolet to mid-infrared wavelength~\cite{Ding2023,Ding2025}. Consequently, the far-infrared (FIR) to millimeter wavelength regime emerges as the sole domain in which accurate measurements of host properties of typical quasars can be made. This regime is characterized by emission from cold interstellar medium (ISM) powered mainly by galaxy-wide star formation~\cite{Leipski2014} except for some rare, and extreme types of active galactic nuclei\cite{Wright2010}. So far, statistical samples of $z \sim 6$ quasars have been observed in both the submillimeter continuum and [CII]158$\mu$m line emission. This enables systematic studies on physical properties of their host galaxies, including sizes, star formation rates, and gas content. In particular, the size of the quasar hosts provides important information on their evolutionary sequence in the context of massive galaxy formation, given the distinct size distribution between star-forming and quiescent galaxies over cosmic time. 

We collect data for all quasars at $5.5 < z < 8$ with available [CII]158$\mu$m observations in the Atacama Large Millimeter Array (ALMA) archive (Methods). All data is consistently reduced using the Common Astronomy Software Applications (CASA) package. Both line fluxes and the 158 $\mu$m continuum are measured using unified methods (Methods 3). 
Our primary sample includes 39 quasars that have both high signal-to-noise ratio [CII] line detection and reliable BH mass measurements. We further remove 9 sources with close companions or complicated sub-structures in their [CII] line maps, indicating possible mergers/interactions. We fit the surface brightness profile of the [CII] line emission for all the remaining 30 sources with a Spergel function in the visibility plane~(Methods 5).
This allows us to derive a more reasonable Sérsic half-light radius ($R_{\rm e}$) compared to the usual Gaussian fitting~\cite{Tan2024}. We exclude 8 targets whose size fitting results could not converge or with large uncertainty, leading to a final sample of 22 quasars.

\begin{figure}
    \centering
    \includegraphics[width=\textwidth]{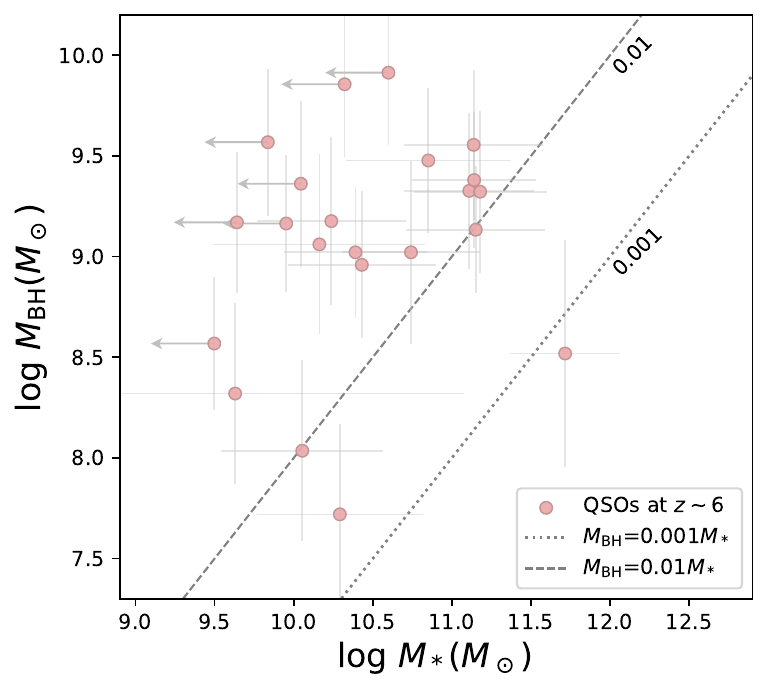}
    \caption{\small{\textbf{The $ \mbh  $-$ \mstar $ distribution of the quasar sample (22 sources).}  The circles are our quasars with reliable size measurements and $\mbh$. The dashed grey line represents $ \mbh  $/$ \mstar =0.01$, and the dotted grey line represents $ \mbh  $/$ \mstar =0.001$. For some sources with grey arrows, their $ \mstar $ are upper limits ($M_{\rm dyn}/2$, assuming that their gas masses are at least half of the $M_{\rm dyn}$). 
    }}
    \label{fig:BH-M*}
\end{figure}

We adopt consistent methods to derive physical parameters for this quasar sample (Methods 6). In particular, dynamical mass ($M_{\rm dyn}$) is derived from the line width and spatial extent of [CII] assuming dynamical equilibrium, and gas mass ($M_{\rm gas}$) is estimated from [CII] luminosity ($L_{\rm [CII]}$) using the calibrated relation in Ref~\cite{Salvestrini2025}. Assuming that the dark matter content is negligible in the inner region of galaxies, we calculate stellar mass ($\mstar$) by subtracting $M_{\rm gas}$ from $M_{\rm dyn}$. 
Figure~\ref{fig:BH-M*} shows the $ \mbh  - \mstar $ relation of our sample, suggesting that these quasars with reliable size measurements generally host overmassive BHs with $\Gamma = \mbh/\mstar \gtrsim 0.01$. This is consistent with many previous studies on these $z \sim 6$ quasars based on their $\mbh$-dynamical mass ratios~\cite{Decarli2018}.

\begin{figure}
    \centering
    \includegraphics[width=\textwidth]{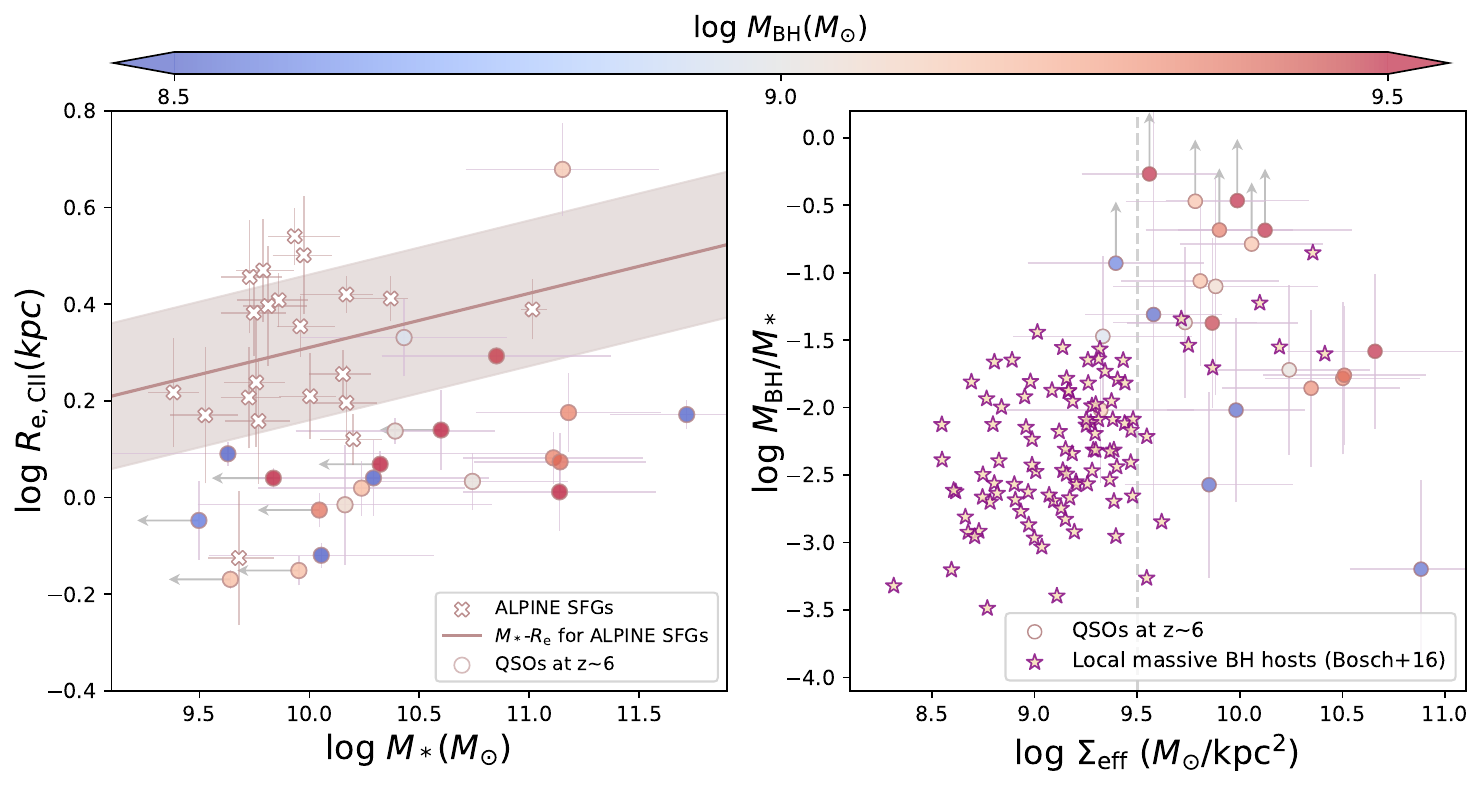}
    \caption{\small{\textbf{Comparison between high-z quasar host galaxies and normal galaxies in terms of galaxy compactness. Left: Comparison between high-z quasars and SFGs in the $R_{\rm e}$-$ \mstar $ plane. Right: Comparison between high-z quasars and local galaxies in the $ \mbh  $/$ \mstar $-$\Sigma_{\rm eff}$ plane.} The circles are our quasars, the open brown crosses are the non-merger SFGs at $z\sim 4-6$ from ALPINE survey~\cite{Fujimoto2020,Faisst2020}, and the purple stars are the local sample selected from literature~\cite{Bosch2016}. Our quasar sample in the two panels are color-coded by $ \mbh  $. The brown solid line and the light brown shaded region are the fitted $M_*-R_{\rm e}$ relation for ALPINE samples and its scatter, respectively. The grey dashed line stands for the $\Sigma_{\rm eff,crit}$ above which the stellar feedback is negligible~\cite{Hopkins2022,Dekel2023,Grudic2018}. For some sources with grey arrows, their $ \mstar $ are upper limits ($M_{\rm dyn}/2$, assuming that their gas masses are at least half of the $M_{\rm dyn}$).
    }}
    \label{fig:BH-compact}
\end{figure}

We first compare the distribution of these quasar hosts in the $\mstar - R_{\rm e}$ plane with that of a representative sample of main-sequence star-forming galaxies (SFGs) at similar redshifts (Figure~\ref{fig:BH-compact}). The sizes of these SFGs, which are mainly from the ALPINE and CRISTAL survey~\cite{Fujimoto2020,Faisst2020,Ikeda2025}, were also measured based on their [CII] line emission with Sérsic profiles (Methods 2.2). 
Nearly all of the quasar hosts in our sample fall below the 1$\sigma$ region of the best-fit $\mstar - R_{\rm e}$ relation for the SFG sample. The median $R_{\rm e,[CII]}$ of the quasar sample and SFG sample are 1.58 kpc and 2.26 kpc, respectively. Considering their similar $\mstar$ distribution, this suggests that these quasar hosts are more compact than normal SFGs. We further verify that this conclusion holds independently of the methods of size fitting and $\mstar$ estimation.


Motivated by this connection between the overmassive BHs and the compactness of their host galaxies, we directly explore the relation between the mass surface density ($\Sigma_{\rm eff}=M_{\rm dyn}/\pi R_{\rm e,[CII]}^2$) and $\Gamma$ for these high-z quasar hosts, and compare them with local galaxies that have accurate $\mbh$ measurements (Methods 2.3). Despite the large redshift differences, a similar trend is found for local galaxies with higher compactness for galaxies hosting overmassive BHs (Figure~\ref{fig:BH-compact}). In particular, the compactness of these $z \sim 6$ quasar hosts is comparable to local galaxies hosting the most overmassive BHs with similar $\Gamma$, suggesting a possible evolutionary link between the two populations (see Method 8 for more discussion).
 
\begin{figure}
    \centering
    \includegraphics[width=\textwidth]{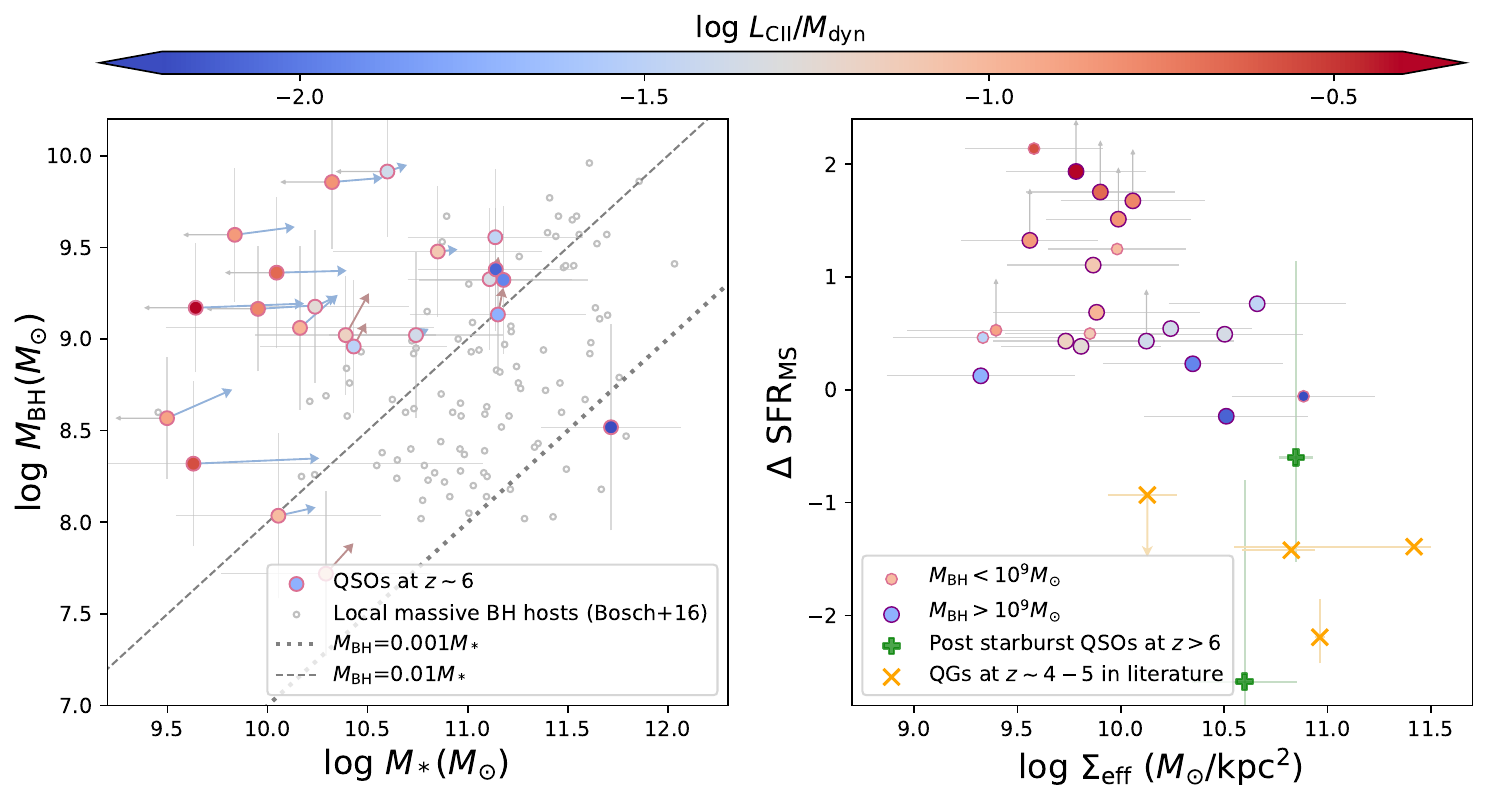}
    \caption{\small{\textbf{Evolution in BH-to-galaxy mass ratio and SFR of quasar hosts. Left: Evolution path of each quasar host in the $ \mbh  $-$ \mstar$ plane.} The circles color-coded by $L_{\rm [CII]}/M_{\rm dyn}$ are our quasar sample, and the open grey dots are the local sample selected from literature~\cite{Bosch2016}. The arrows are the predicted evolution paths, of which blue ones represent galaxy-dominated growth, while light brown ones represent BH-dominated growth. The dashed grey line represents $ \mbh  $/$ \mstar =0.01$, and the dotted grey line represents $ \mbh  $/$ \mstar =0.001$. \textbf{Right: $\Delta \rm SFR_{\rm MS}$-$\Sigma_{\rm e}$ distribution of high-z quasars and quiescent galaxies.} 
    The bigger circles are our quasar sample with $ \mbh  >10^9M_{\odot}$, while smaller circles are those with $ \mbh  <10^9M_{\odot}$. They are color-coded by $L_{\rm [CII]}/M_{\rm dyn}$. For some sources with grey arrows, their $ \mstar $ are upper limits ($M_{\rm dyn}/2$, assuming that their gas masses are at least half of the $M_{\rm dyn}$). The green plus signs are two post-starburst quasar hosts from Ref\cite{Onoue2024,Ding2025}. The yellow crosses are quiescent galaxies at $z\sim 4-5$ with size measurements in the literature~\cite{Carnall2023,Ito2024,Tanaka2024,Graaff2025,Setton2024}.
    }}
    \label{fig:evo-quenching}
\end{figure}

Next we explore how these quasar hosts will evolve and what kind of galaxies will likely be their descendants. To this aim, we compare the growth rates of BHs and SFRs of their host galaxies to probe their evolutionary track in the $ \mbh  - \mstar $ plane. The majority of the quasar sample exhibit elevated BH growth with $\dot{M}_{\rm BH}/{\rm SFR} \sim 0.033$, compared to $\sim$0.005 for local galaxies assuming their growth follows the local $\mbh/\mstar$ relation~\cite{Kormendy2013}. However, this elevated $\dot{M}_{\rm BH}$ does not necessarily mean that those overmassive BHs could maintain their current high $\Gamma$. 
Assuming BHs and their host galaxies could maintain their current growth rate for another gas depletion timescale ($t_{\rm depl}=M_{\rm gas}/$SFR),
we can predict their evolution in the $ \mbh  - \mstar $ plane (Figure~\ref{fig:evo-quenching}).
Despite their elevated BH growth compared to the local $\mbh/\mstar$ relation, these overmassive BH host galaxies cannot maintain their current $\Gamma$ and will move closer to the local relation~\cite{ZhuangM:2023}. This suggests that the growth of host galaxies is delayed compared to their BHs for these overmassive BH hosts, and they should have reached their current high $\Gamma$ by even more rapid BH growth in the past.

With their high stellar masses and SFRs, these overmassive BH hosts are or will soon become the most massive galaxies in the early Universe, which makes them promising progenitors of the first populations of massive quiescent (and compact) galaxies at $z \sim 4-5$~\cite{Xie2024,Carnall2023,Ito2024,Graaff2025}. 
Recent studies show that $\mbh$ is strongly anti-correlated with the cool gas content in massive galaxies~\cite{WangT:2024}, suggesting that these quasar hosts at $z \sim 6$ might not be able to maintain their high gas fraction and SFR for a long time. To put them in the context of massive galaxy transformation, we examine their distribution in the $\Delta$SFR$_{\rm MS}$(=log SFR/$\rm SFR_{\rm MS}$)-$\Sigma_{\rm eff}$ plane in Figure~\ref{fig:evo-quenching}. We employ $L_{\rm [CII]}/M_{\rm dyn}$ as an approximation of the cool gas content ($f_{\rm cool}=M_{\rm gas}/M_{\rm dyn}$). Generally, $f_{\rm cool}$ increases with the starburstness ($\Delta$SFR$_{\rm MS}$), indicating a similar correspondence between cool gas content and star formation as in normal SFGs. Moreover, the most compact quasar hosts tend to have the lowest starburstness and gas fraction. Their compactness is already comparable to the first population of quiescent galaxies formed at $z \sim 4-5$, strongly indicating that these compact quasar hosts are in a transition phase into quiescent galaxies. In addition, these compact and low-$f_{\rm cool}$ galaxies also tend to have the highest $\mbh$ and $\mstar$ among the quasar sample, and show negligible further mass growth in the $ \mbh - \mstar $ plane.
All these features suggest that within the high-z quasar sample, those most compact and low-$f_{\rm cool}$ galaxies are likely at the late stage of their rapid BH and galaxy growth, and will soon be quenched.
This is consistent with recent findings of two quasar host galaxies at $z \sim 6$ exhibiting post-starburst stellar features, of which, one is already quenched and the other is transitioning to quiescence\cite{Onoue2024}. The two post-starburst quasar hosts exhibit similar $\mbh$, $\mstar$, hence $\Gamma$, and stellar mass densities  to our compact low-$f_{\rm cool}$ quasar hosts. Their stellar mass densities are also proposed to be similar to $z\sim 4-5$ quiescent galaxies in COSMOS-web\cite{Ding2025}, strongly supporting the evolutionary connection between our compact and low-$f_{\rm cool}$ galaxies and the $z \sim 4-5$ quiescent galaxies.

These findings provide novel insights into the physical origin of massive BHs hosted by high-z quasars and their connections with the formation and evolution of massive galaxies. An intimate connection between the compactness of host galaxies and formation of overmassive BHs in the early Universe is established. The underlying physics is likely that the high gas densities and deep gravitational potential enable both enhanced gas inflow~\cite{Di2023} and suppressed stellar feedback powered gas expulsion~\cite{Hopkins2022,Dekel2025}, which can effectively fuel the rapid BH growth. The same scenario could also be applicable to another important population of AGNs at $z \sim 4-8$, the ``Little Red Dots" (LRDs), which have been recently discovered by JWST observations. They are generally considered to contain BHs with $\mbh \sim 10^{5-7} M_{\odot}$, which are significantly above the values expected from the local $\mbh-\mstar$ relation. Their point-like morphologies indicate that their host galaxies, if they have, must be extremely compact ($\lesssim 10^2$ pc\cite{Baggen2023,Furtak2023,Chen2025}).

 Together with recent findings on the fundamental role of the SMBHs in governing the gas content and star formation in galaxies in the nearby Universe~\cite{WangT:2024,Martin2018}, a unified picture emerges that $z \sim 6$ quasars with the most massive BHs will likely form the first population of massive quiescent galaxies in the Universe. Depending on their subsequent evolution, they could either maintain their relatively high $\Gamma$ and compactness if they stay untouched, becoming today's compact quiescent galaxies with overmassive BHs, or they could become more normal, and extended quiescent galaxies with lower $\Gamma$ if they experience subsequent major or minor mergers with galaxies hosting less overmassive BHs.


\clearpage

\begin{figure}
    \centering
    \includegraphics[width=\textwidth]{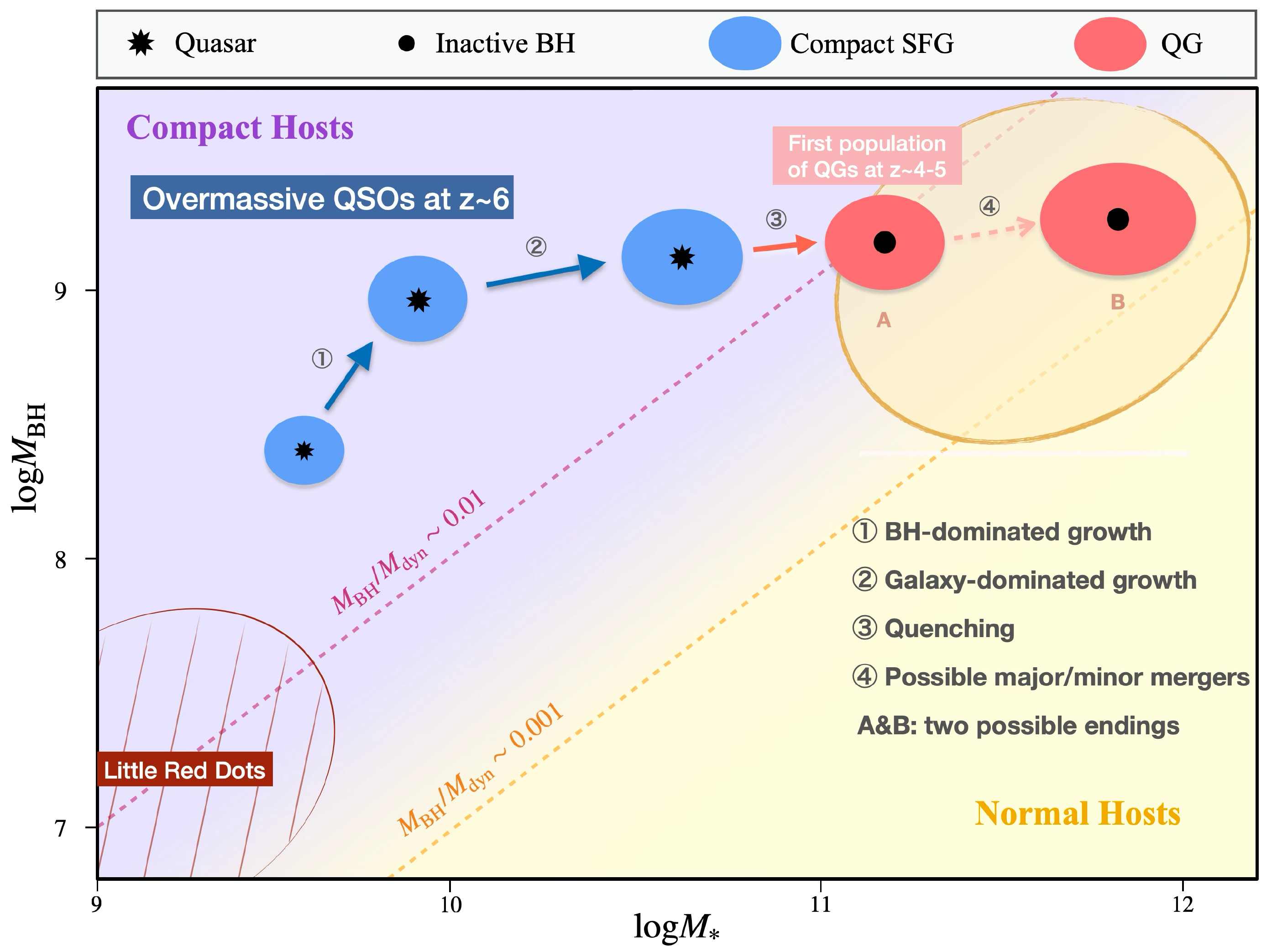}
    \caption{\small{\textbf{Schematic diagram of evolution scenario for overmassive quasar host galaxies.} The yellow dashed line represents $\Gamma \sim 0.001$, while the purple dashed line corresponds to $\Gamma \sim 0.01$. Blue ellipses denote star-forming galaxies, and red ellipses represent quiescent galaxies. We utilize the light purple region and light yellow region to distinguish between compact and normal galaxies, respectively. The evolution path, depicted by two dark blue arrows and one red arrow, illustrates the most recent or ongoing two growth stages and the subsequent quenching of overmassive quasars. Notably, the starting point of this path does not indicate the formation point of these overmassive BHs. We show possible positions that these quasars would reach in the nearby Universe with the translucent orange region. Within this region, we use two red ellipses to illustrate two representative possibilities. The one labeled by A represents that they stay undisturbed and become local compact overmassive BH hosts, while the one labeled by B represents that they undergo subsequent mergers and become normal elliptical galaxies in the local universe. 
    The red shaded region shows the location of Little Red Dots (LRDs) which also host overmassive BHs and are compact discovered by JWST. It is highly likely that they would also exhibit a similar growth pattern.
    }}
    \label{fig:evolution}
\end{figure}

\clearpage
\begin{methods}
\subsection{1. Cosmology:}
We adopt a Kroupa IMF~\cite{Kroupa2001} to estimate star formation rates, and a $\rm \Lambda$CDM cosmology with $H_{0}$ = 70 km s$^{-1}$ Mpc$^{-1}$, $\Omega_{M}$ = 0.3, and $\Omega_{\Lambda}$ = 0.7 throughout this paper. 
\subsection{2. Sample selection:}
\subsection{2.1 The quasar sample} 
We assemble an initial sample of 97 quasars at $5.5<z<8$ from archival ALMA [CII] observations and reduce them uniformly (Methods 3). After excluding 11 sources lacking [CII] line detections, 2 sources without direct BH mass or bolometric luminosity measurements, and 6 sources with resolved close companions within 2.5'' to avoid possible interactions, we measure the [CII] flux for the remaining 78 sources using aperture photometry (Methods 4). To ensure the robust size measurements via Spergel profile fitting in GILDAS (Methods 5), we only retain 35 sources with [CII] flux signal-to-noise ratios (S/N$_{F_{\rm [CII]}}$) above 20~\cite{Tan2024}. During the size fitting, we further exclude 5 targets with complicated (sub-)structures (e.g., unresolved close companions or merging systems) that introduce unreliability to their size, BH mass, and/or [CII] luminosity measurements. Therefore, our primary sample contains 30 targets, among which 3 are flagged due to non-convergence and another 5 are flagged due to low relative accuracy ($R_{\rm e}/\sigma(R_{\rm e})<3$). The final quasar sample comprises 22 quasars. The basic properties of all 86 sources with detected [CII] line are listed in Extended Data Table~\ref{etb1}.

\subsection{2.2 The high-z SFG sample} Based on the ALPINE and CRISTAL surveys, we select those with [CII] size and stellar mass measurements. Specifically, their sizes are measured with Sérsic profiles with the Sérsic index fixed to 1 (exponential-disk profile), which is claimed to be suitable for these SFGs~\cite{Fujimoto2020}, and their stellar masses are measured by SED fitting~\cite{Faisst2020}. We further exclude galaxies classified as pair mergers or those with relative accuracy of sizes ($R_{\rm e}/\sigma(R_{\rm e})$) below 3 to ensure the reliability of their sizes. Therefore, the final high-z SFG sample contains 21 galaxies with redshifts ranging from 4.41 to 5.67.

\subsection{2.3 The local massive non-active BH hosts sample}
Based on the local sample from Ref ~\cite{Bosch2016} which has direct $ \mbh  $ measurement, 
we select galaxies whose $ \mstar $ are higher than $10^{9.5} M_{\odot}$ to focus on the massive local galaxies, and exclude those whose $ \mbh  $ are upper limits and(or) measured through reverberation mapping to ensure the reliability. Consequently, the local sample comprises 166 galaxies in total. The stellar mass ($ \mstar $) is derived through $ \mstar /L_{\rm K}$=0.10$\sigma_{\rm e}^{0.45}$ and $R_{\rm e}$ is the K-band effective radius. 
 

\subsection{3. Reduction of ALMA data for the quasar sample}

For projects in ALMA Cycles higher than 0, we restore the calibrated visibilities through running the default scripts in their corresponding versions of CASA (from 4.2.0 to 6.1.1) and split each target out. Then for each target, main procedures of data reduction and the details of each step are listed as follows:

\begin{enumerate}
    \item[3.1] We re-compute and set the weights for observations in cycle 0-2 with the CASA task {\sc statwt}. We correct the phase centers through the CASA task {\sc fixvis} and coordinate reference systems of different observations through directly modification of the {\sc field} and {\sc source} tables to make them consistent. Then we use the CASA task {\sc concat} to combine all measurement sets (MSs) belonging to the same target to one MS, preparing for the subsequent imaging. 
    \item[3.2] Data cube containing both line emission and continuum: All spectral windows (SPWs) covering the [C II] line are chosen to be imaged using the CASA task {\sc tclean}. We set the specmode to {\sc cube}, deconvolver to Högbom, width to 2 ($\sim 31.25$ MHz), threshold to 2$\sigma$ ($\sigma$ is the rms noise) and weighting to Natural algorithm to achieve maximum signal-to-noise ratio (S/N).
    \item[3.3] Line spectrum with continuum: We extract the full spectrum from an aperture whose radius is manually chosen ($\sim$ 0.3-1.3") based both on the beam size and the extent of the source shown in the moment zero map of line channels. Then we fit the spectrum with a single Gaussian plus constant profile to determine the [C II] redshift and line width (FWHM). 
    \item[3.4] [CII] intensity map: Based on the [C II] properties, we define the spectral range which contains 90\% of the total line flux to be line channels, that is $\pm 0.7\times$FWHM from the line center. Line-free channels are defined to be $\pm 1.5 \times$FWHM away from the line center. By fitting the line-free channels using the CASA task {\sc imcontsub} with {\sc fitorder}=1, we create a continuum-subtracted data cube. The [C II] intensity map is produced by collapsing all line channels with the CASA task {\sc immoments} (moment 0). 
    \item[3.5] Continuum map: Based on line-free channels in all SPWs, we create the continuum map using the CASA task {\sc tclean} with specmode set to {\sc mfs}, deconvolver to Högbom, threshold to 2$\sigma$ ($\sigma$ is the rms noise) and weighting to Natural algorithm. 
\end{enumerate}

\subsection{4. Aperture flux measurement:}
\begin{enumerate}
    \item [4.1] [CII] line flux: 
    Considering the possibly various beam sizes across line channels introduced by different projects for some targets, we first convert flux densities in each channel map of the line cube from Jy beam$^{-1}$ to Jy pixel$^{-1}$ through dividing by the beam area in pixels. Then channels with beam size variations $<0.01$'' are grouped into subsets. Each subset is collapsed into an integrated intensity map on which we perform aperture photometry. To determine the optimal aperture, we test circular apertures (0.1–1.5'' radii, 0.1'' steps). For each aperture size, we: 1) measure the source flux within the testing aperture centered on the intensity peak; 2) randomly sample 5000 background fluxes using the same testing aperture within a 2–5'' annulus centered on the target; and 3) fit a Gaussian distribution to these fluxes to obtain its standard deviation as the RMS noise. We select the aperture radius that maximizes the S/N$_{F_{\rm [CII]}}$ and apply aperture corrections to account for flux outside the aperture. Therefore, the measured total flux of our defined line channels is the sum of corrected fluxes from all sub-maps for targets with beam variations, or for others, is just this corrected flux. Lastly, since the defined line channels only cover 90\% of the total line flux, we further divide the measured total flux by 0.9 to convert to the total line flux.
    \item [4.2] Continuum flux density: We perform aperture photometry directly on the continuum map, following similar steps to determine the optimal aperture radius and apply aperture correction.
    
    

\end{enumerate}


\subsection{5. Size measurement:} 
\begin{enumerate}
\item[5.1]\textbf{2D-Gaussian fitting in CASA:} We quickly measure the deconvolved size of the [C II] emission by fitting a 2D Gaussian to the source in its [CII] line map using the CASA task {\sc imfit}.  

\item[5.2]\textbf{Spergel fitting in GILDAS:}
Note that the deconvolved size provided by the {\sc imfit} is the FWHM of the major axis of a 2D Gaussian profile. However, the general surface brightness profile of a galaxy is normally described as a Sérsic function. Therefore, we turn to Spergel profile which closely approximates the Sérsic profile and can be analytically transformable into Fourier space, enabling direct fitting in the uv-plane. The Spergel model has been implemented in {\sc \verb|uv_fit|} in GILDAS and has been verified to perform effectively ~\cite{Tan2024}.
\begin{enumerate}
    \item[5.2.1]\textbf{Production of single-channel visibility for [C II]:} We generate the single-channel visibilities for line emissions to serve as input files for \verb|uv_fit| in GILDAS. For the line emission, we first subtract continuum emission in the uv-plane with {\sc uvcontsub} and then average the line channels in each SPW of the initial MS into a single-channel visibility via {\sc split}. After correcting the channel width of each visibility to be consistent, we use {\sc concat} to combine them into one SPW by setting a large enough {\sc freqtol}(e.g., 1GHz). The one-channel uv data are exported into uvfits files with {\sc exportuvfits} prepared for analysis in GILDAS.
    \item[5.2.2] \textbf{Initial value of parameters:} The {\sc \verb|e_spergel|} model has seven parameters. Their initial values are calculated as follows: 1) \verb|xoff| and \verb|yoff|: the position offsets in R.A. and Dec are calculated by subtracting the position of the phase center from the peak x and y positions provided by {\sc imfit}; 2) \verb|flux|: it is obtained by converting the flux provided by {\sc imfit} to Jansky; 3) \verb|maj| and \verb|min|: the effective radius of major and minor axis are approximated as 1/2.43~\cite{Murphy2017} of the FWHM major and minor axis provided by {\sc imfit}; 4) \verb|pa|: it is the same with the position angle from {\sc imfit}; 5) \verb|nu|: the Spergel index are initially set to 0.5 (i.e., n=1 in Sérsic profile). 
    \item[5.2.3] \textbf{Tuning of parameters:} We update input values of model parameters with the results from each fitting iteration. When the Spergel index \verb|nu| stabilizes (variation $<\pm 0.01$ between iterations), we fix its value (within $-1 - 0.5$ where -1 is the theoretical limit~\cite{Tan2024}) to reduce the degrees of freedom, making the size measurements more robust, and then re-run the {\sc \verb|uv_fit|} to achieve final results with improved relative accuracy. Targets with non-convergent fits or low-confidence major axis measurements (\verb|maj|/$\sigma$(\verb|maj|)$<3$) are flagged as unsuccessful. In the Extended Figure~\ref{fig:demo_of_gildas}, PJ359-06 and J2304+0045 are shown as examples for quasars with successful fitting and other sources in our primary sample with convergent but low relative accuracy fittings respectively. The profile of normalized visibility amplitude of J2304+0045 reveals a sharp decrease at the long uv distance end, likely driving its $R_{\rm e,maj}/\sigma (R_{\rm e,maj})<3$. The elevated reduced-$\chi^2$ for J2304+0045 reflects systematic residuals in the visibility profile, also explaining its lower measurement confidence.

     \item[5.2.4] \textbf{Conversion of sizes fitted from Spergel to Sérsic model:} As proposed in Ref~\cite{Tan2024}, the effective radius estimated by the Spergel model is systematically smaller than when using the Sérsic model due to the difference in their indices. Therefore, based on the conversion function (Eq. 4 in Ref~\cite{Tan2024}), we calculate the ratio $R_{\rm e, \text{Sérsic}}/R_{\rm e, \text{Spergel}}$ to obtain the corrected Sérsic size, which could then be compared to the sizes of the high-z SFG sample.
\end{enumerate}





\end{enumerate}

\subsection{6. Derivation of basic properties:}
We collect BH mass measurements and absolute magnitude $M_{1450}$ of our quasar host galaxies from Ref~\cite{Wang2013,Willot2013,DeRosa2014,Willot2015,Wu2015,Willot2017,Decarli2018,Izumi2018,Izumi2019,Reed2019,Shen2019,Andika2020,Eilers2020,Izumi2021a,Neeleman2021,Izumi2021b,Eilers2020,Izumi2021a,Yang2021,Izumi2021b}. Caveats of the assumptions used during the derivation 
are briefly illustrated within each sub-section. See Methods 7 for a more comprehensive and detailed discussion regarding their potential influences on our primary conclusions.
\begin{enumerate}
    \item[6.1] \textbf{Continuum luminosity and star formation rates} The dust continuum of each sample mainly probes observed-frame 1.2 mm or rest-frame 158 $\mu$m, which is close to the peak of the far-infrared SED, allowing a rather accurate estimate on their $L_{\rm IR}$. Therefore, we scale the commonly used modified blackbody model with dust temperature $T_{\rm dust} = 47$K and emissivity index $\beta=1.6$ ~\cite{Wang2013,Decarli2018,Venemans2018,Venemans2019,Venemans2020} to the continuum flux density we measured, and then integrate the model within 8-1000 $\mu$m to derive the total infrared (TIR) luminosity ($L_{\rm TIR}$).
    Assuming that all the cold dust emission is heated by star formation, we obtain the star formation rate from the $L_{\rm TIR}$ through SFR$_{\rm IR} = 1.48\times10^{-10} L_{\rm TIR}$~\cite{Murphy2011}. Note that the SFR derived here suffers from uncertainties introduced by our assumptions on dust properties ($T_{\rm dust}$ and $\beta$) and the assumed no AGN contribution to the cold dust emission. 

    \item[6.2] \textbf{[C II] source sizes} Based on the Sérsic effective radii of [CII] emission on major and minor axes ($R_{\rm e,maj}$,$R_{\rm e,min}$) obtained in Method 4, we derive the circularized effective radius $R_{\rm e}=\sqrt{R_{\rm e,maj}R_{\rm e,min}}$.
 

    \item[6.3] \textbf{Dynamical masses} Based on the virial theorem, the dynamical mass within effective radius could be estimated by $M_{\rm dyn} = V_{\rm circ}^{2} R_{\rm e}/G$. Different kinetic models provide various estimates of $V_{\rm circ}$. Here we assume a widely adopted rotating thin disk model~\cite{Wang2013,Willot2015,Willot2017,Venemans2016,Izumi2018,Izumi2019}
    \begin{equation}
      M_{\rm dyn} = 1.16\times10^{5}(0.75{\rm FWHM}_{\rm [CII]}/{\rm sin}i)^{2}D,
      \label{wang}
    \end{equation}
     where D is the diameter set to 2$R_{\rm e,[CII]}$, $i$ is the inclination angle of the thin disk (cos$i$=b/a, where b/a is the [C II] minor and major axis ratio). As for the uncertainty of $M_{\rm dyn}$, we consider the measurement error in FWHM and size, along with a systematic uncertainty of 0.3 dex.
     Note that for galaxies whose velocities are dominated by non-order rotation or whose orderly rotating disks are thick, the $M_{\rm dyn}$ derived here might be overestimated to different extents. Nevertheless, this discrepancy is acceptable as it leads to underestimated $\Gamma$, suggesting that our selection of overmassive BHs still stands and could be even conservative (see Methods 7 for more discussions).

    \item[6.4] \textbf{Line luminosity and gas mass} The [CII] luminosity is derived from the line flux measured by 2D Gaussian fitting, using 
    \begin{equation}
         L_{[\rm{CII}]} / L_{\odot}=1.04 \times 10^{-3} \nu_{[\rm{CII}], \text { obs }} F_{[\rm{CII}]} D_L^2,
    \end{equation}
   where $\nu_{[\rm{CII}], \text { obs }}$ is the observed frequency of the [CII] line in GHz, and $D_L$ is the luminosity distance in Mpc. Then we convert the $L_{\rm{[CII]}}$ to molecular gas mass using a scaling relation $\log \left(M_{\rm gas} / M_{\odot}\right)=0.75_{-0.31}^{+0.31} \log L_{\rm{[CII]}}+2.87_{-0.07}^{+0.07}$ derived from some quasars at $z>7$~\cite{Salvestrini2025}, with an intrinsic dispersion of 0.1 dex. We consider this intrinsic dispersion as systematic uncertainty for $M_{\rm gas}$.
   
   \item[6.5] \textbf{Stellar mass} We primarily estimate the $ \mstar $ by subtracting $M_{\rm gas}$ from $M_{\rm dyn}$, assuming that the dark matter content of these galaxies could be negligible in the inner few kpc. Systematic uncertainties in both $M_{\rm gas}$ and $M_{\rm dyn}$ are taken into account when deriving the uncertainty of $ \mstar $ (u\_$ \mstar $). For most galaxies, their $ \mbh  $/$ \mstar  \sim 0-0.1$. However, in some cases, $M_{\rm gas}$ are too high, resulting in $ \mbh  $/$ \mstar $ exceeding 0.1 or even falling below 0 (i.e. $ \mstar <0$). Given that $M_{\rm gas}$ has been found to be comparable to $ \mstar $ in these systems~\cite{Neeleman2021,Decarli2022,Kaasinen2024}, we use half of the $M_{\rm dyn}$ as upper limits for $ \mstar $ of these galaxies (shown with grey arrows in figures).
   \item[6.6] \textbf{SFR offset to the MS ($\Delta \rm SFR_{\rm MS}$)} 
Based on the $ \mstar $, we derive the star-forming main sequence for each quasar with
\begin{equation}
\begin{aligned}
\log _{10} \mathrm{SFR}_{\mathrm{MS}}= & 0.84 \times (\log _{10} \mstar -10) -(0.15+0.026 \times (\log _{10} \mstar  -10)) \times t_{\text {cosmic age }}+1.89.
\end{aligned}
  \label{gas mass}
\end{equation}
Therefore, $\Delta \rm SFR_{\rm MS} \equiv \log _{10}(\mathrm{SFR} / \mathrm{SFR}_{\mathrm{MS}})$. For quasars whose $ \mstar $ are upper limits, their $\Delta \rm SFR_{\rm MS}$ given by this method are lower limits.

    \item[6.7] \textbf{Bolometric luminosity, BH mass and BH growth rates} We convert the absolute magnitude $M_{1450}$ to $L_{1450}$, and from which derive the $L_{\rm bol}$=4.2$L_{1450}$~\cite{Wang2013}. For quasars with Mg II-based $ \mbh  $ collected from literature, considering the uncertainty of single-epoch estimation, we apply a typical systematic uncertainty of 0.3 dex~\cite{Shen2008}. For 7 quasars without $ \mbh  $ but with $L_{\rm bol}$, we derive their $ \mbh $ from $L_{\rm bol}$ by assuming an Eddington ratio of 1. Their $ \mbh  $ uncertainties are given 0.45 dex~\cite{Willot2015}. We test the reasonableness of this Eddington limit assumption by applying it to those already with Mg II-based $ \mbh $, and find a general consistency between the two $ \mbh $ decided by different methods. Then we characterize the growth of BH mass ($\dot{M}_{\rm BH}$) with $L_{\rm bol}$=[$\eta$ /(1-$\eta$)]$\dot{M}_{\rm BH}$c$^{2}$, where $\eta$ is the radiative efficiency assumed to be 0.1, suitable for moderately accreting BHs ($\sim$75\% of our quasars with Mg II-based $\mbh$ exhibit $-1 <{\rm log}\lambda_{\rm Edd} \lesssim 0$).

\end{enumerate}

\subsection{7. Robustness of our main conclusions to variations in primary sample selection, size measurement methods and uncertainties in $M_{\rm dyn}$ and SFR:} 
We discuss the effects of relaxing our sample selection criteria (S/N$_{F_{\rm [CII]}}$ threshold), changing the size fitting profile, and uncertainties in $M_{\rm dyn}$ ane SFR estimations on our main conclusions.

\begin{enumerate}

    \item[7.1] Expanding the primary sample with a less strict requirement on S/N$_{F_{\rm [CII]}}$: The quasar sample, though clean, is limited in size due to our strict S/N$_{F_{\rm [CII]}}$ threshold. To test for selection bias, we determine a lower threshold by analyzing sources without companions or substructures. They are classified into successful fitting group and unsuccessful fitting group. The former requires for convergent fitting results with \verb|maj|/$\sigma$(\verb|maj|)$>3$, while the later is the rest. The Extended Figure~\ref{compare_size_fitting_requirement} directly shows that the unsuccessful fitting group generally have lower $F_{\rm [CII]}$ and S/N$_{F_{\rm [CII]}}$. Notice that the lowest S/N$_{F_{\rm [CII]}}$ for the successful fitting group and the median S/N$_{F_{\rm [CII]}}$ for the unsuccessful fitting group are both around 10, we test to relax the threshold from 20 to 10, expanding the primary sample to 54 targets (35 with reliable size measurements). Extended Figure~\ref{changes_effects} (top)  shows the $\mbh - \mstar$ and $\mstar -R_{\rm e}$ distributions for the expanded quasar sample. Despite the inclusion of lower-S/N$_{F_{\rm [CII]}}$ sources, $\sim$70\% of our quasars retain compact host sizes, confirming the robustness of our size conclusion.

    \item[7.2] Testing to change the size fitting profile to exponential: While the free Spergel index ($\nu$) fitting should be optimal for our quasar sample according to the spheroidal morphology found in compact submm-bright galaxies~\cite{Tan2024b}, we test to use consistent size fitting profile with the high-z SFG sample by fixing $\nu$ to 0.5 (exponential profile). We re-measure sizes for the same quasar sample already selected in this work, and then update their $\mstar$. Their sizes are still smaller than typical SFGs (Extended Figure~\ref{changes_effects}, middle).

    \item[7.3] \textbf{Testing an empirical estimation of $M_{\rm dyn}$}: The turbulent dynamics, high velocity dispersion and thick disks, reported for some high-z quasars~\cite{Neeleman2021,Yue2021} or sub-mm galaxies~\cite{Tan2024b}, could all lead to a lower $V_{\rm circ}$, and thus lower $M_{\rm dyn}$ than our estimation. In the bottom panel of Extended Figure~\ref{changes_effects}, we test the impacts on their BH-to-galaxy mass ratios and mass-size distribution when using an empirical estimate of $V_{\rm circ}\sim$0.5FWHM/sin$i$~\cite{Neeleman2021} which is fitted from a sample containing both dispersion-dominated and ordered rotation-dominated systems. It reveals that with lower $M_{\rm dyn}$, more galaxies would have uncertain $\mstar$ estimation and have generally higher BH-to-galaxy mass ratios. Even in this case, their galaxy sizes remain smaller compared to those of normal SFGs with similar masses, suggesting that our main conclusion still stands. Therefore, we opt to still use the potentially overestimated $M_{\rm dyn}$ to prevent any possible exaggeration of BH-to-galaxy mass ratio.

    \item[7.4] \textbf{Examing the effects of AGN contribution on SFR estimation}: The estimation of SFR largely relies on assumptions concerning dust properties and the fraction of SF contribution to $L_{\rm IR}$. Generally, warmer and/or optically thick dust leads to higher $L_{\rm FIR}$, but a degeneracy exists between dust temperature and AGN contribution~\cite{Shangguan2018}, complicating the estimation of the intrinsic SFR for each individual source. Since the related main conclusion is that some most compact gas-poor galaxies will quench earlier than others, we need to ensure that there exists no systematic difference in AGN contribution to $L_{\rm IR}$ between the two subgroups. This ensures that applying a unified approach to estimate SFRs for the entire sample is appropriate. Therefore, we use $L_{\rm bol}$ to trace the AGN contribution, and compare distributions of $L_{\rm bol}$ for the most compact gas-poor sources and others. These two subgroups appear to follow the same $L_{\rm bol}$ distribution according to the KS test (p-value=0.96). After excluding the possibility of systematic difference between the two subgroups, any re-calibration favoring higher or lower SFR would only systematically alters $\Delta \rm SFR_{\rm MS}$ for the entire sample, while leaving our main conclusion unaffected.

\end{enumerate}

\subsection{8. Connection between some local compact quiescent galaxies hosting overmassive BHs and these high-z overmassive BH hosts:} Over the past decade, several local galaxies such as NGC 1277, NGC 4486B, NGC 1332 and NGC 1271 have been found to host overmassive BHs ($ \mbh  /M_{\rm bulge}>$ 5\%, while the expected one is 0.3\%~\cite{Kormendy2013}), posing a question on how to explain the formation of these outliers. Previous research has proposed two possible explanations: (1) they are stripped galactic nuclei, making central BHs used to be normal become relatively overmassive~\cite{Seth2014}; (2) they are the undisturbed relics of earlier Universe ($z>2$) where $ \mbh  /M_{\rm bulge}$ is supposed to be higher~\cite{FerreMateu2015}. The former favours less massive galaxies which are not isolated, while the latter expects galaxies to be compact and have stellar populations older than 10 Gyr. Both of them can successfully explain some examples. However, the uncertainties in BH mass estimation stemming from different modeling approaches make it more complicated and controversial, as some of these BHs could have an order of magnitude lower mass (e.g., NGC 1277). 
In this work, we find that compared to overmassive BH hosts in the local universe (hereafter, local outliers), those at $z \sim 6$ (hereafter, high-z outliers) have a slightly higher $ \mbh  / \mstar $ which will slightly decline during subsequent short-period galaxy-dominated growth. Additionally, high-z outliers also exhibit compactness comparable to local outliers (despite not tracing stars currently). But assuming that all their gas and dust will finally form stars in situ, and if those local outliers do exist, we propose that, at least for part of the compact local outliers with undisturbed morphologies, some high-z outliers that could remain undisturbed until $z\sim 0$ might be their promising progenitors.

\end{methods}

\newpage

\setcounter{figure}{0}
\renewcommand{\figurename}{Extended Data Figure}
\begin{Extended Data Figure}[!ht]
	\centering
 \includegraphics[width=\textwidth]{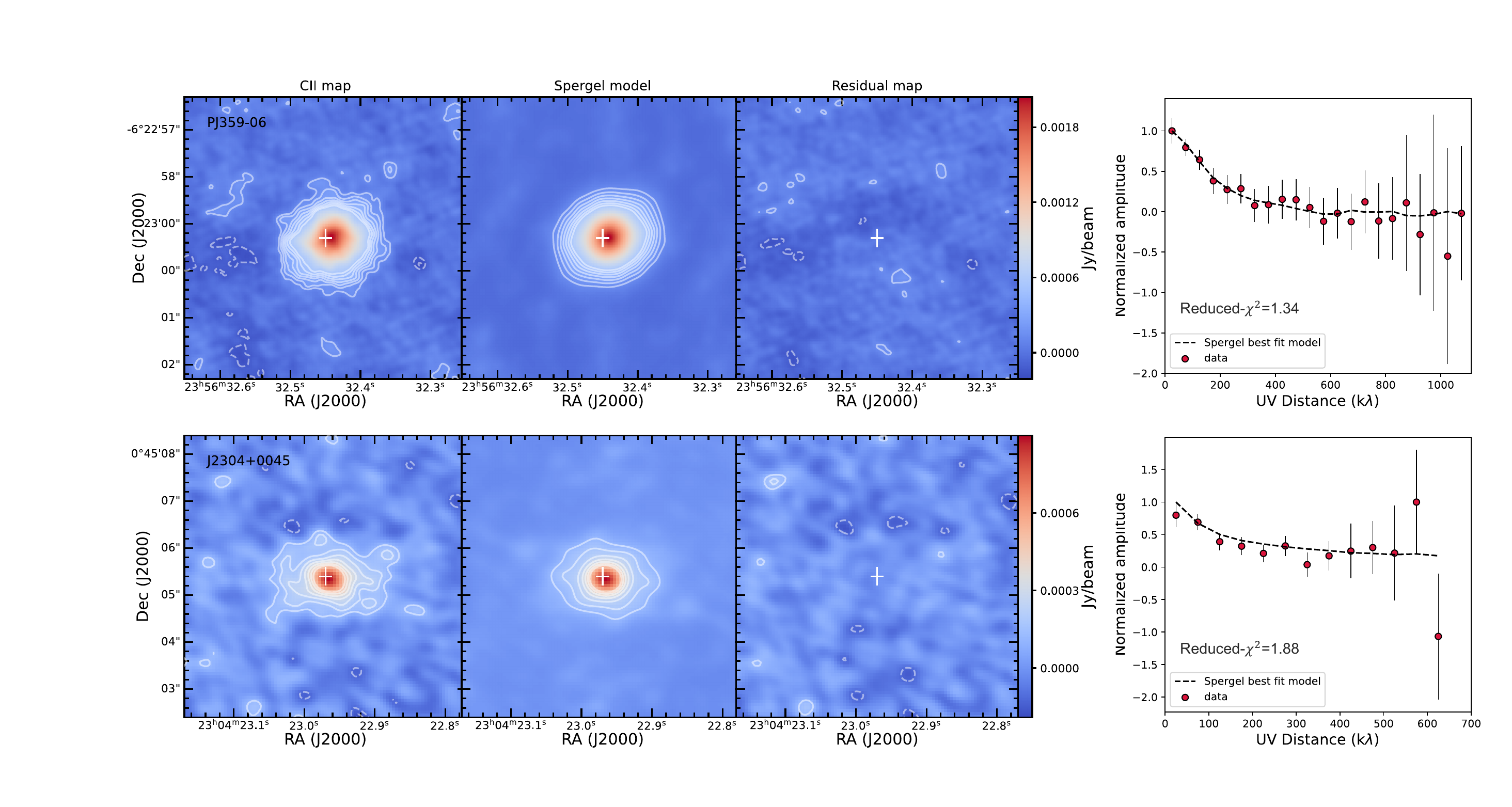}
    \caption{\small{\textbf{Two representative examples to show our size fitting results for successful and unsuccessful measurements.} The two representative sources are selected out due to showing median S/N$_{F_{\rm [CII]}}$  from the samples with and without successful size measurements respectively, where the unsuccessful fitting samples are sources whose fits converge but have $R_{\rm e}$/$\sigma$($R_{\rm e}$)$<3$. We show the dirty images produced with Natural weighting, the source models we fit convolved with the dirty beam, the residual maps after subtracting the source model and the normalized visibility amplitude as a function of uv distance for the two sources, respectively. Contours in these images start at $\pm 3\sigma$ and increase by a factor of 1.5. The white crosses represent their phase centers.}}
    \label{fig:demo_of_gildas}
\end{Extended Data Figure}

\begin{Extended Data Figure}[!ht]
	\centering
	\includegraphics[width=0.7\textwidth]{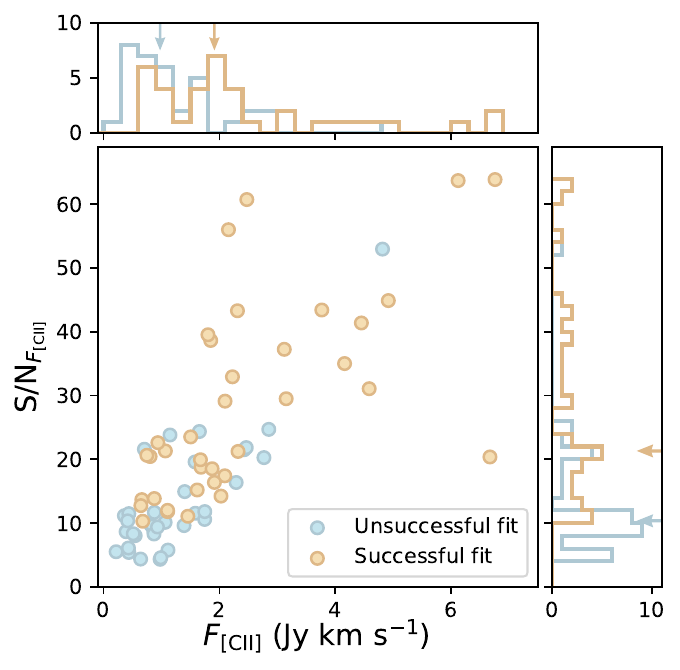}
    \caption{\small{\textbf{Comparison between successful and unsuccessful fitted sources in respect of $F_{\rm [CII]}$ and S/N$_{F_{\rm [CII]}}$.} All the dots and open bars are sources without close companions or complicated sub-structures. The blue ones are 35 unsuccessfully fitted sources whose results are nonconvergent or with $R_{\rm e}$/$\sigma$($R_{\rm e}$)$<3$, while the yellow ones are 37 sources successfully fitted. The blue and yellow arrows in the two histograms show the median values for the two samples.}}
    \label{compare_size_fitting_requirement}
\end{Extended Data Figure}

\begin{Extended Data Figure}[!ht]
	\centering
	\includegraphics[width=0.8\textwidth]{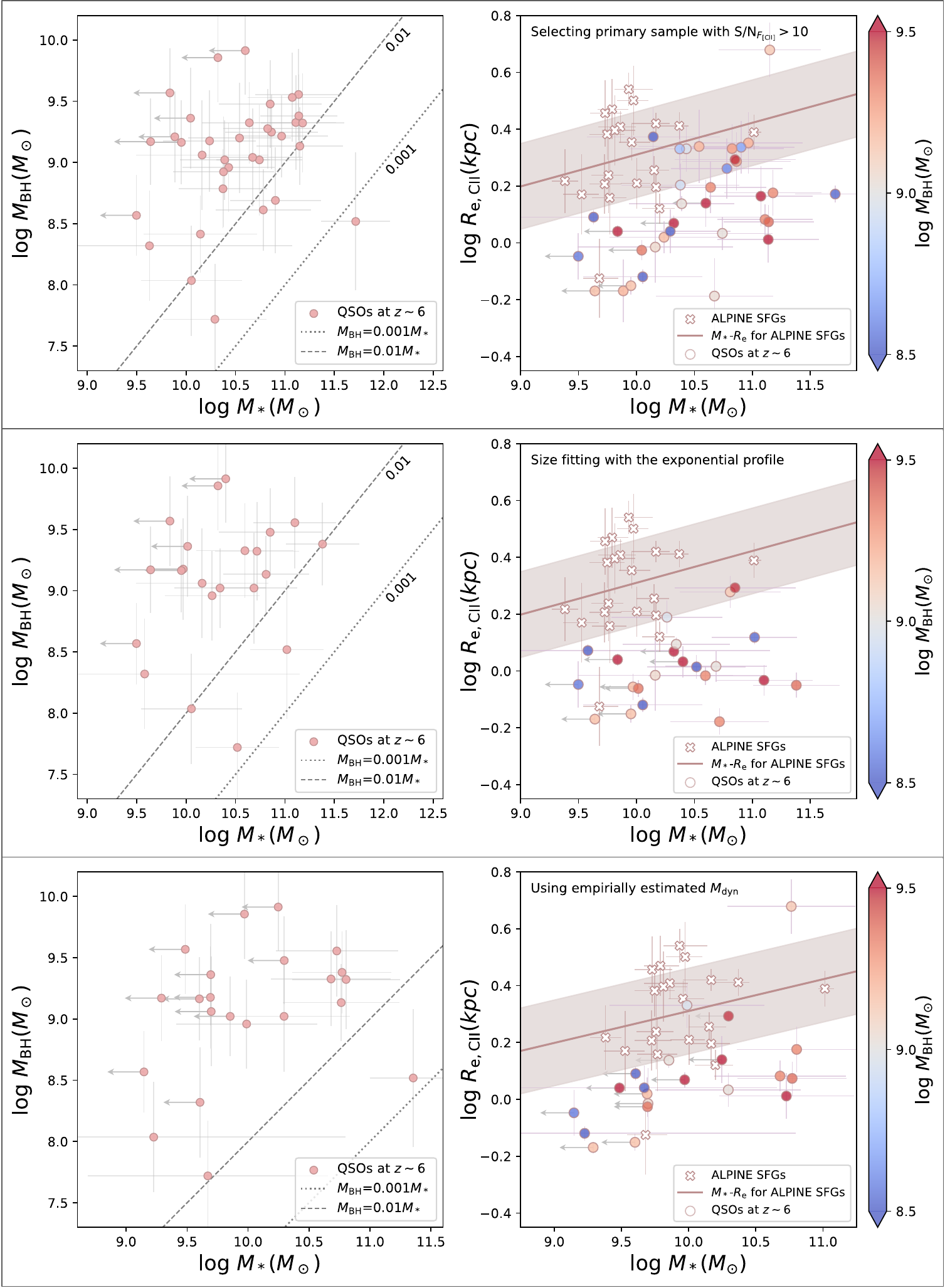}
    \caption{\small{\textbf{Test the robustness of our main conclusion related to galaxy sizes when reducing the required S/N$_{F_{\rm [CII]}}$ threshold (top), and changing the size fitting profile (middle) and $M_{\rm dyn}$ estimation method (bottom).} The pink dots in the left figure and the color-coded circles in the right figure of each panel are the quasar sample with reliable size measurements (only changed in the top panel).}}
    \label{changes_effects}
\end{Extended Data Figure}



\newpage
\onecolumn
\noindent\textbf{\large Extended Data}
\renewcommand\thefigure{\arabic{figure}}
\setcounter{figure}{0}
\renewcommand{\tablename}{Extended Data Table}
\begin{small}
\begin{ThreePartTable}
\setlength{\tabcolsep}{3pt}
\setlength\LTleft{-0.6in} 
\LTcapwidth=\textwidth
\begin{longtable}{ccccccccc}
\caption{\textbf{Basic properties of 86 sources with detected [CII] line}}
\label{etb1}\\
\vspace{-1.cm}\\
\hline \hline 
Name & Project ID & z & log $R_{\rm e}$ & log $L_{\rm [CII]}$ & log $L_{\rm IR}$ & log $L_{\rm bol}$ & log $M_{\rm BH}$ & log $M_{\rm dyn}$ \\ 
~ & ~ & ~ & (log kpc) & (log $L_{\odot}$) & (log $L_{\odot}$) & (log $L_{\odot}$) & (log $M_{\odot}$) & (log $M_{\odot}$) \\ 
\midrule
\vspace{-1.1cm}
\endfirsthead\\
\multicolumn{9}{c}{(continued)}\\
\hline \hline 
Name & Project ID & z & log $R_{\rm e}$ & log $L_{\rm [CII]}$ & log $L_{\rm IR}$ & log $L_{\rm bol}$ & log $M_{\rm BH}$ & log $M_{\rm dyn}$ \\ 
~ & ~ & ~ & (log kpc) & (log $L_{\odot}$) & (log $L_{\odot}$) & (log $L_{\odot}$) & (log $M_{\odot}$) & (log $M_{\odot}$) \\ 

\midrule
\vspace{-0.7cm}
\endhead\\
J0038-1527  & \makecell{2018.1.01188.S} & 7.0335 & $ 0.68  \pm 0.10 $          & $9.44 \pm 0.02$ & $12.67 \pm 0.02 $ & $13.85 \pm 0.09 $ & $9.13 \pm 0.32$          & $11.18 \pm 0.41 $ \\
J0055+0146  & \makecell{2012.1.00676.S} & 6.0055 & $       ...      $\tnote{a} & $8.31 \pm 0.08$ & $11.74 \pm 0.07 $ & $12.92 \pm 0.09 $ & $8.38 \pm 0.36$          & $19.46 \pm 86.25$ \\ 
J0100+2802  & \makecell{2015.1.00692.S} & 6.3266 & $       ...      $\tnote{a} & $9.46 \pm 0.02$ & $12.44 \pm 0.13 $ & $14.70 \pm 0.08 $ & $9.99 \pm 0.31$          & $15.80 \pm 0.77 $ \\
J0109-3047  & \makecell{2012.1.00882.S\\2013.1.00273.S\\2015.1.00399.S} & 6.7904 & $ 0.02  \pm 0.06 $          & $9.08 \pm 0.02$ & $12.18 \pm 0.02 $ & $13.20 \pm 0.06 $ & $9.18 \pm 0.42$          & $10.34 \pm 0.37 $ \\
J0129-0035  & \makecell{2011.0.00206.S\\2012.1.00240.S\\2015.1.00997.S} & 5.7788 & $ -0.12 \pm 0.03 $          & $9.22 \pm 0.01$ & $12.864 \pm 0.004$ & $12.55 \pm 0.09 $ & $8.04 \pm 0.45$\tnote{*} & $10.24 \pm 0.33 $ \\
J0136+0226  & \makecell{2019.1.00074.S} & 6.2129 & $ -0.05 \pm 0.08 $          & $8.90 \pm 0.02$ & $11.70 \pm 0.07 $ & $12.75 \pm 0.01 $ & $8.57 \pm 0.33$          & $9.80  \pm 0.40 $ \\
J0142-3327  & \makecell{2015.1.01115.S\\2017.1.01301.S} & 6.3374 & $ 0.94  \pm 0.23 $          & $9.69 \pm 0.01$ & $12.97 \pm 0.01 $ & $14.12 \pm 0.09 $ & $9.60 \pm 0.45$\tnote{*} & $11.39 \pm 0.53 $ \\
J0210-0456  & \makecell{2011.0.00243.S} & 6.4328 & $       ...      $\tnote{a} & $8.61 \pm 0.05$ & $11.77 \pm 0.07 $ & $12.78 \pm 0.09 $ & $7.90 \pm 0.43$          & $10.11 \pm 0.50 $ \\
J0218+0007  & \makecell{2018.1.01188.S} & 6.7695 & $ -0.65 \pm 0.23 $          & $9.08 \pm 0.04$ & $12.60 \pm 0.01 $ & $13.21 \pm 0.09 $ & $8.79 \pm 0.35$          & $9.94  \pm 0.57 $ \\
J0224-4711  & \makecell{2015.1.00606.S} & 6.5220 & $ 0.08  \pm 0.05 $          & $9.81 \pm 0.01$ & $12.99 \pm 0.01 $ & $13.77 \pm 0.02 $ & $9.33 \pm 0.39$          & $11.16 \pm 0.36 $ \\
J0227-0605  & \makecell{2019.1.00074.S} & 6.2100 & $ 1.40  \pm 0.44 $\tnote{b} & $8.68 \pm 0.04$ & $11.64 \pm 0.11 $ & $      ...      $ & $8.34 \pm 0.31$          & $11.32 \pm 0.76 $ \\
J0229-0808  & \makecell{2019.1.01025.S} & 6.7248 & $ 0.11  \pm 0.03 $          & $9.66 \pm 0.01$ & $13.394 \pm 0.004$ & $      ...      $ & $     ...     $          & $10.56 \pm 0.35 $ \\
J0244-5008  & \makecell{2017.1.01472.S} & 6.7308 & $ -0.01 \pm 0.12 $          & $9.37 \pm 0.01$ & $12.42 \pm 0.02 $ & $13.68 \pm 0.02 $ & $9.06 \pm 0.45$          & $10.35 \pm 0.43 $ \\
J0246-5219  & \makecell{2019.1.01025.S} & 6.8871 & $ 0.03  \pm 0.06 $          & $9.42 \pm 0.01$ & $12.75 \pm 0.01 $ & $13.14 \pm 0.09 $ & $9.02 \pm 0.45$          & $10.81 \pm 0.37 $ \\
J0252-0503  & \makecell{2019.1.01025.S} & 7.0013 & $ 1.86  \pm 0.25 $          & $9.01 \pm 0.05$ & $12.57 \pm 0.02 $ & $13.65 \pm 0.09 $ & $9.11 \pm 0.33$          & $12.17 \pm 0.60 $ \\
J0305-3150  & \makecell{2017.1.01532.S} & 6.6139 & $ 0.18  \pm 0.03 $\tnote{c} & $9.64 \pm 0.19$ & $13.19 \pm 0.09 $ & $13.38 \pm 0.02 $ & $8.98 \pm 0.34$          & $11.11 \pm 0.36 $ \\
J0313-1806  & \makecell{2019.1.01025.S} & 7.6416 & $       ...      $\tnote{a} & $8.86 \pm 0.05$ & $12.40 \pm 0.02 $ & $13.45 \pm 0.09 $ & $9.21 \pm 0.41$          & $9.52  \pm 0.56 $ \\
J0319-1008  & \makecell{2018.1.01188.S} & 6.8272 & $ -0.37 \pm 0.15 $          & $9.04 \pm 0.10$ & $12.26 \pm 0.04 $ & $13.14 \pm 0.09 $ & $8.60 \pm 0.33$          & $10.44 \pm 0.57 $ \\
J0411-0907  & \makecell{2019.1.01025.S} & 6.8255 & $ 0.78  \pm 0.12 $\tnote{c} & $8.95 \pm 0.05$ & $11.94 \pm 0.06 $ & $13.63 \pm 0.09 $ & $8.98 \pm 0.34$          & $11.36 \pm 0.49 $ \\
J0430-1445  & \makecell{2018.1.01188.S} & 6.7136 & $ 0.18  \pm 0.13 $          & $9.35 \pm 0.03$ & $12.36 \pm 0.03 $ & $      ...      $ & $     ...     $          & $10.41 \pm 0.45 $ \\
J0454-4448  & \makecell{2015.1.01115.S} & 6.0585 & $       ...      $\tnote{a} & $8.79 \pm 0.10$ & $12.28 \pm 0.03 $ & $13.56 \pm 0.09 $ & $9.04 \pm 0.45$\tnote{*} & $10.47 \pm 0.65 $ \\
J0525-2406  & \makecell{2019.1.01025.S} & 6.5393 & $ 0.21  \pm 0.02 $\tnote{b} & $9.70 \pm 0.01$ & $13.19 \pm 0.01 $ & $13.18 \pm 0.09 $ & $8.46 \pm 0.36$          & $10.23 \pm 0.34 $ \\
J0706+2921  & \makecell{2018.1.01188.S} & 6.6036 & $ 0.20  \pm 0.09 $          & $9.20 \pm 0.04$ & $12.46 \pm 0.02 $ & $13.97 \pm 0.09 $ & $9.32 \pm 0.33$          & $10.69 \pm 0.42 $ \\
J0842+1218  & \makecell{2016.1.00544.S} & 6.0753 & $ 0.35  \pm 0.11 $          & $8.80 \pm 0.03$ & $12.14 \pm 0.02 $ & $13.76 \pm 0.09 $ & $9.21 \pm 0.34$          & $10.98 \pm 0.45 $ \\
J0859+0022  & \makecell{2016.1.01423.S} & 6.3900 & $       ...      $\tnote{a} & $8.57 \pm 0.04$ & $11.62 \pm 0.04 $ & $12.63 \pm 0.04 $ & $7.30 \pm 0.33$          & $10.97 \pm 0.46 $ \\
J0909+0440  & \makecell{2019.1.00074.S} & 6.1293 & $ 0.53  \pm 0.23 $          & $9.13 \pm 0.03$ & $12.59 \pm 0.01 $ & $12.95 \pm 0.09 $ & $8.44 \pm 0.45$\tnote{*} & $10.32 \pm 0.55 $ \\
J0910+1656  & \makecell{2018.1.01188.S} & 6.7288 & $ 0.26  \pm 0.11 $          & $9.09 \pm 0.04$ & $12.01 \pm 0.05 $ & $13.13 \pm 0.09 $ & $8.61 \pm 0.33$          & $10.81 \pm 0.45 $ \\
J0910-0414  & \makecell{2018.1.01188.S} & 6.6361 & $ 0.01  \pm 0.08 $          & $9.70 \pm 0.01$ & $13.291 \pm 0.003$ & $13.64 \pm 0.09 $ & $9.56 \pm 0.37$          & $11.18 \pm 0.40 $ \\
J0921+0007  & \makecell{2018.1.01188.S} & 6.5643 & $ 0.37  \pm 0.10 $          & $8.97 \pm 0.03$ & $11.86 \pm 0.08 $ & $13.07 \pm 0.09 $ & $8.41 \pm 0.32$          & $10.25 \pm 0.45 $ \\
J0923+0402  & \makecell{2018.1.01188.S} & 6.6331 & $ 0.29  \pm 0.10 $          & $9.26 \pm 0.02$ & $12.06 \pm 0.05 $ & $13.67 \pm 0.09 $ & $9.25 \pm 0.30$          & $10.90 \pm 0.43 $ \\
J0923+0753  & \makecell{2019.1.01025.S} & 6.6814 & $ 0.34  \pm 0.09 $          & $8.87 \pm 0.04$ & $11.92 \pm 0.05 $ & $13.19 \pm 0.09 $ & $8.69 \pm 0.43$          & $10.92 \pm 0.44 $ \\
J1007+2115  & \makecell{2019.1.01025.S} & 7.5151 & $ -0.03 \pm 0.15 $          & $9.50 \pm 0.02$ & $12.85 \pm 0.01 $ & $13.69 \pm 0.09 $ & $9.16 \pm 0.37$          & $10.36 \pm 0.46 $ \\
J1044-0125  & \makecell{2011.0.00206.S\\2012.1.00240.S\\2015.1.00997.S} & 5.7854 & $ 0.16  \pm 0.11 $          & $9.16 \pm 0.03$ & $12.94 \pm 0.01 $ & $13.95 \pm 0.09 $ & $9.53 \pm 0.36$          & $11.09 \pm 0.43 $ \\
J1048-0109  & \makecell{2015.1.01115.S\\2017.1.01301.S} & 6.6755 & $ -0.03 \pm 0.04 $          & $9.69 \pm 0.01$ & $13.387 \pm 0.004$ & $13.39 \pm 0.09 $ & $9.36 \pm 0.41$          & $10.35 \pm 0.35 $ \\
J1058+2930  & \makecell{2019.1.01025.S} & 6.5849 & $ 0.26  \pm 0.19 $          & $9.18 \pm 0.05$ & $12.22 \pm 0.05 $ & $13.27 \pm 0.09 $ & $8.73 \pm 0.32$          & $10.81 \pm 0.53 $ \\
J1104+2134  & \makecell{2018.1.01188.S} & 6.7661 & $ 0.17  \pm 0.33 $          & $9.25 \pm 0.04$ & $13.098 \pm 0.005$ & $13.65 \pm 0.09 $ & $9.23 \pm 0.34$          & $11.55 \pm 0.64 $ \\
J1120+0641  & \makecell{2012.1.00882.S} & 7.0848 & $ 0.07  \pm 0.06 $          & $9.05 \pm 0.02$ & $12.31 \pm 0.01 $ & $13.63 \pm 0.04 $ & $9.38 \pm 0.34$          & $11.15 \pm 0.38 $ \\
J1137+0045  & \makecell{2019.1.00074.S} & 6.3672 & $ 0.14  \pm 0.08 $          & $9.51 \pm 0.01$ & $12.51 \pm 0.02 $ & $13.750 \pm 0.002 $ & $9.91 \pm 0.36$          & $10.90 \pm 0.39 $ \\
J1146+0124  & \makecell{2019.1.00074.S} & 6.2498 & $ 0.41  \pm 0.04 $\tnote{b} & $9.26 \pm 0.01$ & $12.24 \pm 0.02 $ & $12.47 \pm 0.09 $ & $7.96 \pm 0.45$\tnote{*} & $10.60 \pm 0.35 $ \\
J1152+0055  & \makecell{2015.1.01115.S\\2016.1.01423.S} & 6.3639 & $ 0.47  \pm 0.31 $          & $8.72 \pm 0.05$ & $11.64 \pm 0.07 $ & $13.05 \pm 0.02 $ & $8.53 \pm 0.45$\tnote{*} & $10.89 \pm 0.62 $ \\
J1202-0057  & \makecell{2016.1.01423.S} & 5.9294 & $       ...      $\tnote{a} & $8.81 \pm 0.02$ & $11.81 \pm 0.02 $ & $12.13 \pm 0.03 $ & $7.61 \pm 0.45$\tnote{*} & $10.52 \pm 0.54 $ \\
J1205-0000  & \makecell{2019.1.00074.S} & 6.7225 & $ 0.53  \pm 0.25 $          & $9.26 \pm 0.02$ & $12.83 \pm 0.01 $ & $12.81 \pm 0.11 $ & $9.34 \pm 0.46$          & $11.36 \pm 0.56 $ \\
J1207+0630  & \makecell{2015.1.01115.S} & 6.0361 & $ 0.44  \pm 0.21 $\tnote{b} & $9.02 \pm 0.07$ & $12.09 \pm 0.05 $ & $13.65 \pm 0.09 $ & $9.13 \pm 0.45$\tnote{*} & $11.50 \pm 0.57 $ \\
J1208-0200  & \makecell{2017.1.00541.S} & 6.1162 & $ 0.12  \pm 0.23 $          & $8.63 \pm 0.04$ & $11.56 \pm 0.10 $ & $12.89 \pm 0.01 $ & $8.85 \pm 0.53$          & $10.11 \pm 0.55 $ \\
J1217+0131  & \makecell{2019.1.00074.S} & 6.2089 & $ 0.60  \pm 0.88 $          & $8.91 \pm 0.04$ & $11.75 \pm 0.09 $ & $13.15 \pm 0.09 $ & $8.64 \pm 0.45$\tnote{*} & $11.58 \pm 1.19 $ \\
J1243+0100  & \makecell{2019.1.00074.S} & 7.0749 & $ 0.17  \pm 0.03 $          & $9.47 \pm 0.01$ & $12.956 \pm 0.004$ & $12.65 \pm 0.03 $ & $8.52 \pm 0.56$          & $11.72 \pm 0.34 $ \\
J1306+0356  & \makecell{2015.1.01115.S\\2017.1.01301.S} & 6.0331 & $ 0.35  \pm 0.05 $\tnote{b} & $9.46 \pm 0.02$ & $12.74 \pm 0.01 $ & $13.72 \pm 0.09 $ & $9.32 \pm 0.31$          & $10.35 \pm 0.37 $ \\
J1319+0950  & \makecell{2011.0.00206.S\\2012.1.00240.S} & 6.1341 & $ 0.33  \pm 0.09 $          & $9.31 \pm 0.02$ & $12.85 \pm 0.01 $ & $13.81 \pm 0.09 $ & $9.28 \pm 0.33$          & $10.87 \pm 0.40 $ \\
J1342+0928  & \makecell{2017.1.00396.S} & 7.5415 & $ 0.33  \pm 0.08 $          & $8.99 \pm 0.02$ & $12.43 \pm 0.02 $ & $13.70 \pm 0.09 $ & $8.96 \pm 0.36$          & $10.49 \pm 0.41 $ \\
J1406-0116  & \makecell{2019.1.00074.S} & 6.2976 & $ 0.09  \pm 0.03 $          & $9.70 \pm 0.01$ & $13.423 \pm 0.004$ & $12.83 \pm 0.09 $ & $8.32 \pm 0.45$\tnote{*} & $10.26 \pm 0.33 $ \\
J1509-1749  & \makecell{2015.1.01115.S} & 6.1222 & $ 1.24  \pm 0.35 $          & $9.23 \pm 0.04$ & $12.63 \pm 0.01 $ & $13.85 \pm 0.09 $ & $9.34 \pm 0.45$\tnote{*} & $12.15 \pm 0.66 $ \\
J2002-3013  & \makecell{2019.1.01025.S} & 6.6875 & $ -0.17 \pm 0.11 $          & $9.32 \pm 0.02$ & $12.98 \pm 0.01 $ & $13.75 \pm 0.09 $ & $9.21 \pm 0.37$          & $10.19 \pm 0.43 $ \\
J2054-0005  & \makecell{2018.1.00908.S} & 6.0391 & $ -0.17 \pm 0.02 $          & $9.55 \pm 0.01$ & $13.223 \pm 0.003$ & $13.48 \pm 0.09 $ & $9.17 \pm 0.35$          & $9.94  \pm 0.34 $ \\
J2100-1715  & \makecell{2015.1.01115.S\\2017.1.01301.S} & 6.0811 & $ 0.07  \pm 0.19 $          & $9.22 \pm 0.04$ & $12.46 \pm 0.03 $ & $13.21 \pm 0.09 $ & $9.65 \pm 0.42$          & $10.74 \pm 0.50 $ \\
J2102-1458  & \makecell{2018.1.01188.S} & 6.6649 & $ 0.02  \pm 0.18 $          & $9.24 \pm 0.02$ & $12.65 \pm 0.01 $ & $13.21 \pm 0.09 $ & $8.87 \pm 0.36$          & $10.04 \pm 0.49 $ \\
J2211-3206  & \makecell{2015.1.01115.S} & 6.3384 & $       ...      $\tnote{a} & $8.65 \pm 0.08$ & $12.24 \pm 0.03 $ & $13.68 \pm 0.09 $ & $9.16 \pm 0.45$\tnote{*} & $8.73  \pm 0.72 $ \\
J2211-6320  & \makecell{2019.1.01025.S} & 6.8445 & $       ...      $\tnote{a} & $9.45 \pm 0.02$ & $12.92 \pm 0.01 $ & $13.15 \pm 0.09 $ & $8.74 \pm 0.49$          & $10.61 \pm 0.55 $ \\
J2216-0016  & \makecell{2016.1.01423.S} & 6.0963 & $ 0.33  \pm 0.07 $          & $8.79 \pm 0.03$ & $11.47 \pm 0.07 $ & $12.52 \pm 0.02 $ & $8.79 \pm 0.32$          & $10.42 \pm 0.40 $ \\
J2228+0152  & \makecell{2017.1.00541.S} & 6.0807 & $ 0.72  \pm 0.38 $          & $8.60 \pm 0.04$ & $11.34 \pm 0.13 $ & $12.59 \pm 0.02 $ & $8.08 \pm 0.45$\tnote{*} & $10.84 \pm 0.70 $ \\
J2229+1457  & \makecell{2012.1.00676.S} & 6.1508 & $       ...      $\tnote{a} & $8.92 \pm 0.05$ & $11.56 \pm 0.10 $ & $12.91 \pm 0.09 $ & $8.08 \pm 0.40$          & $9.32  \pm 0.60 $ \\
J2239+0207  & \makecell{2017.1.00541.S} & 6.2499 & $ -0.19 \pm 0.13 $          & $9.22 \pm 0.02$ & $12.78 \pm 0.01 $ & $12.87 \pm 0.02 $ & $9.04 \pm 0.40$          & $10.73 \pm 0.45 $ \\
J2304+0045  & \makecell{2019.1.00074.S} & 6.3502 & $ 1.04  \pm 0.21 $          & $9.07 \pm 0.02$ & $12.58 \pm 0.01 $ & $12.71 \pm 0.01 $ & $8.19 \pm 0.45$\tnote{*} & $11.30 \pm 0.51 $ \\
J2310+1855  & \makecell{2018.1.00597.S} & 6.0027 & $ 0.07  \pm 0.03 $          & $9.80 \pm 0.02$ & $13.355 \pm 0.004$ & $14.11 \pm 0.09 $ & $9.86 \pm 0.36$          & $10.62 \pm 0.34 $ \\
J2318-3029  & \makecell{2015.1.01115.S\\2018.1.00908.S} & 6.1455 & $ -0.15 \pm 0.03 $          & $9.48 \pm 0.01$ & $13.219 \pm 0.003$ & $13.48 \pm 0.09 $ & $9.16 \pm 0.34$          & $10.25 \pm 0.34 $ \\
J2318-3113  & \makecell{2015.1.01115.S\\2017.1.01301.S} & 6.4430 & $ 0.20  \pm 0.07 $          & $9.30 \pm 0.03$ & $12.24 \pm 0.04 $ & $13.44 \pm 0.09 $ & $8.92 \pm 0.45$\tnote{*} & $10.49 \pm 0.39 $ \\
J2348-3054  & \makecell{2013.1.00273.S\\2015.1.00399.S} & 6.9010 & $ 0.18  \pm 0.08 $          & $9.24 \pm 0.02$ & $12.799 \pm 0.005$ & $13.28 \pm 0.06 $ & $9.32 \pm 0.40$          & $11.20 \pm 0.40 $ \\
PJ004+17    & \makecell{2017.1.00332.S} & 5.8170 & $       ...      $\tnote{a} & $8.89 \pm 0.05$ & $12.21 \pm 0.02 $ & $13.40 \pm 0.09 $ & $8.88 \pm 0.45$\tnote{*} & $11.42 \pm 0.57 $ \\
PJ007+04    & \makecell{2015.1.01115.S\\2017.1.01301.S} & 6.0011 & $ 1.35  \pm 0.36 $          & $9.42 \pm 0.02$ & $13.016 \pm 0.004$ & $13.65 \pm 0.09 $ & $9.18 \pm 0.41$          & $11.70 \pm 0.66 $ \\
PJ009-10    & \makecell{2015.1.01115.S\\2017.1.01301.S} & 6.0028 & $ 0.41  \pm 0.03 $\tnote{c} & $9.60 \pm 0.02$ & $12.83 \pm 0.01 $ & $13.61 \pm 0.09 $ & $9.38 \pm 0.40$          & $11.06 \pm 0.37 $ \\
PJ011+09    & \makecell{2017.1.00332.S} & 6.4702 & $ 0.75  \pm 0.24 $          & $9.07 \pm 0.08$ & $12.64 \pm 0.01 $ & $13.73 \pm 0.09 $ & $9.14 \pm 0.45$          & $11.76 \pm 0.57 $ \\
PJ036+03    & \makecell{2015.1.00399.S} & 6.5404 & $ 0.04  \pm 0.02 $          & $9.28 \pm 0.01$ & $12.785 \pm 0.004$ & $13.93 \pm 0.09 $ & $9.57 \pm 0.36$          & $10.14 \pm 0.33 $ \\
PJ056-16    & \makecell{2017.1.00332.S} & 5.9665 & $ 0.25  \pm 0.40 $          & $8.60 \pm 0.07$ & $11.46 \pm 0.10 $ & $13.68 \pm 0.09 $ & $8.85 \pm 0.32$          & $11.17 \pm 0.73 $ \\
PJ065-19    & \makecell{2015.1.01115.S} & 6.1251 & $       ...      $\tnote{a} & $8.96 \pm 0.05$ & $12.08 \pm 0.04 $ & $13.64 \pm 0.09 $ & $9.13 \pm 0.45$\tnote{*} & $10.79 \pm 0.50 $ \\
PJ065-26    & \makecell{2015.1.01115.S\\2017.1.01301.S} & 6.1872 & $ 0.85  \pm 0.16 $          & $9.35 \pm 0.03$ & $12.71 \pm 0.01 $ & $13.89 \pm 0.09 $ & $9.66 \pm 0.35$          & $11.89 \pm 0.47 $ \\
PJ083+11    & \makecell{2019.1.01436.S} & 6.3401 & $ 0.55  \pm 0.01 $\tnote{c} & $9.58 \pm 0.01$ & $13.180 \pm 0.003$ & $13.662 \pm 0.002$ & $9.30 \pm 0.42$          & $10.44 \pm 0.33 $ \\
PJ158-14    & \makecell{2017.1.00332.S} & 6.0686 & $ 0.19  \pm 0.06 $\tnote{c} & $9.82 \pm 0.01$ & $12.997 \pm 0.003$ & $13.96 \pm 0.09 $ & $9.20 \pm 0.33$          & $11.14 \pm 0.36 $ \\
PJ159-02    & \makecell{2015.1.01115.S} & 6.3814 & $ 0.34  \pm 0.13 $          & $9.06 \pm 0.04$ & $12.24 \pm 0.03 $ & $13.71 \pm 0.09 $ & $9.20 \pm 0.45$\tnote{*} & $10.59 \pm 0.46 $ \\
PJ167-13    & \makecell{2015.1.00606.S\\2015.1.01115.S\\2016.1.0544.S} & 6.5152 & $ 0.46  \pm 0.03 $\tnote{c} & $9.52 \pm 0.02$ & $12.36 \pm 0.01 $ & $13.24 \pm 0.09 $ & $8.48 \pm 0.44$          & $11.17 \pm 0.34 $ \\
PJ183+05    & \makecell{2015.1.01115.S\\2016.1.00544.S} & 6.4387 & $ 0.29  \pm 0.02 $          & $9.85 \pm 0.01$ & $13.393 \pm 0.004$ & $13.41 \pm 0.09 $ & $9.48 \pm 0.36$          & $10.95 \pm 0.42 $ \\
PJ217-16    & \makecell{2015.1.01115.S} & 6.1482 & $       ...      $\tnote{a} & $8.98 \pm 0.09$ & $12.05 \pm 0.07 $ & $13.77 \pm 0.09 $ & $9.25 \pm 0.45$\tnote{*} & $11.40 \pm 0.80 $ \\
PJ231-20    & \makecell{2015.1.01115.S\\2016.1.00544.S} & 6.5864 & $ 2.47  \pm 0.45 $\tnote{b} & $9.69 \pm 0.01$ & $13.316 \pm 0.003$ & $13.87 \pm 0.09 $ & $9.61 \pm 0.39$          & $13.04 \pm 0.76 $ \\
PJ239-07    & \makecell{2017.1.00332.S} & 6.1102 & $       ...      $\tnote{a} & $8.93 \pm 0.04$ & $11.79 \pm 0.05 $ & $13.98 \pm 0.09 $ & $9.48 \pm 0.31$          & $16.01 \pm 13.34$ \\ 
PJ308-21    & \makecell{2015.1.01115.S\\2016.A.00018.S} & 6.2339 & $ 0.87  \pm 0.07 $\tnote{c} & $9.50 \pm 0.02$ & $12.49 \pm 0.02 $ & $13.53 \pm 0.09 $ & $9.23 \pm 0.37$          & $11.88 \pm 0.38 $ \\
PJ323+12    & \makecell{2018.1.00908.S} & 6.5873 & $ 0.13  \pm 0.06 $\tnote{c} & $9.14 \pm 0.04$ & $11.94 \pm 0.08 $ & $13.82 \pm 0.09 $ & $9.05 \pm 0.35$          & $10.48 \pm 0.42 $ \\
PJ359-06    & \makecell{2017.1.01301.S} & 6.1720 & $ 0.14  \pm 0.03 $          & $9.34 \pm 0.01$ & $12.33 \pm 0.02 $ & $13.71 \pm 0.09 $ & $9.02 \pm 0.32$          & $10.51 \pm 0.35 $ \\
VIMOS2911   & \makecell{2015.1.00606.S} & 6.1492 & $ 0.04  \pm 0.08 $          & $9.32 \pm 0.01$ & $12.32 \pm 0.01 $ & $12.23 \pm 0.09 $ & $7.72 \pm 0.45$\tnote{*} & $10.43 \pm 0.38 $ \\

\hline 
\end{longtable}
\vspace{-1cm}
\begin{tablenotes}
\footnotesize               
\item[a] The size fitting results could not converge.
\item[b] They have resolved close companions within 2.5'', therefore we use multiple models simultaneously to fit their sizes.
\item[c] They have companions or sub-structures that cannot be visually detected before size fitting, but are revealed in the residual maps. Therefore, their flux measurement and $\mbh$ measurement could be unreliable.
\item[*]  Their $\mbh$ are derived from $L_{\rm bol}$ by assuming an Eddington ratio of 1. The $\mbh$ for others are collected from the literature~\cite{DeRosa2014,Izumi2018,Izumi2019,Reed2019,Shen2019,Andika2020,Eilers2020,Izumi2021a,Neeleman2021,Yang2021,Izumi2021b}

\end{tablenotes}
\end{ThreePartTable}
\end{small}

\bibliography{main.bib}

\begin{thebibliography}{10}
\expandafter\ifx\csname url\endcsname\relax
  \def\url#1{\texttt{#1}}\fi
\expandafter\ifx\csname urlprefix\endcsname\relax\def\urlprefix{URL }\fi
\providecommand{\bibinfo}[2]{#2}
\providecommand{\eprint}[2][]{\url{#2}}

\bibitem{Fan2006}
\bibinfo{author}{{Fan}, X.} \emph{et~al.}
\newblock \bibinfo{title}{{A Survey of z{\ensuremath{>}}5.7 Quasars in the
  Sloan Digital Sky Survey. IV. Discovery of Seven Additional Quasars}}.
\newblock \emph{\bibinfo{journal}{\aj}} \textbf{\bibinfo{volume}{131}},
  \bibinfo{pages}{1203--1209} (\bibinfo{year}{2006}).

\bibitem{Wang2013}
\bibinfo{author}{{Wang}, R.} \emph{et~al.}
\newblock \bibinfo{title}{{Star Formation and Gas Kinematics of Quasar Host
  Galaxies at z {\ensuremath{\sim}} 6: New Insights from ALMA}}.
\newblock \emph{\bibinfo{journal}{\apj}} \textbf{\bibinfo{volume}{773}},
  \bibinfo{pages}{44} (\bibinfo{year}{2013}).

\bibitem{Jiang2016}
\bibinfo{author}{{Jiang}, L.} \emph{et~al.}
\newblock \bibinfo{title}{{The Final SDSS High-redshift Quasar Sample of 52
  Quasars at z{\ensuremath{>}}5.7}}.
\newblock \emph{\bibinfo{journal}{\apj}} \textbf{\bibinfo{volume}{833}},
  \bibinfo{pages}{222} (\bibinfo{year}{2016}).

\bibitem{Decarli2018}
\bibinfo{author}{{Decarli}, R.} \emph{et~al.}
\newblock \bibinfo{title}{{An ALMA [C II] Survey of 27 Quasars at z
  {\ensuremath{>}} 5.94}}.
\newblock \emph{\bibinfo{journal}{\apj}} \textbf{\bibinfo{volume}{854}},
  \bibinfo{pages}{97} (\bibinfo{year}{2018}).

\bibitem{Venemans2020}
\bibinfo{author}{{Venemans}, B.~P.} \emph{et~al.}
\newblock \bibinfo{title}{{Kiloparsec-scale ALMA Imaging of [C II] and Dust
  Continuum Emission of 27 Quasar Host Galaxies at z {\ensuremath{\sim}} 6}}.
\newblock \emph{\bibinfo{journal}{\apj}} \textbf{\bibinfo{volume}{904}},
  \bibinfo{pages}{130} (\bibinfo{year}{2020}).

\bibitem{Reines2015}
\bibinfo{author}{{Reines}, A.~E.} \& \bibinfo{author}{{Volonteri}, M.}
\newblock \bibinfo{title}{{Relations between Central Black Hole Mass and Total
  Galaxy Stellar Mass in the Local Universe}}.
\newblock \emph{\bibinfo{journal}{\apj}} \textbf{\bibinfo{volume}{813}},
  \bibinfo{pages}{82} (\bibinfo{year}{2015}).

\bibitem{Kormendy2013}
\bibinfo{author}{{Kormendy}, J.} \& \bibinfo{author}{{Ho}, L.~C.}
\newblock \bibinfo{title}{{Coevolution (Or Not) of Supermassive Black Holes and
  Host Galaxies}}.
\newblock \emph{\bibinfo{journal}{\araa}} \textbf{\bibinfo{volume}{51}},
  \bibinfo{pages}{511--653} (\bibinfo{year}{2013}).

\bibitem{Ding2023}
\bibinfo{author}{{Ding}, X.} \emph{et~al.}
\newblock \bibinfo{title}{{Detection of stellar light from quasar host galaxies
  at redshifts above 6}}.
\newblock \emph{\bibinfo{journal}{\nat}} \textbf{\bibinfo{volume}{621}},
  \bibinfo{pages}{51--55} (\bibinfo{year}{2023}).

\bibitem{Ding2025}
\bibinfo{author}{{Ding}, X.} \emph{et~al.}
\newblock \bibinfo{title}{{SHELLQs-JWST Unveils the Host Galaxies of Twelve
  Quasars at $z{\ensuremath{>}}6$}}.
\newblock \emph{\bibinfo{journal}{arXiv e-prints}}
  \bibinfo{pages}{arXiv:2505.03876} (\bibinfo{year}{2025}).

\bibitem{Leipski2014}
\bibinfo{author}{{Leipski}, C.} \emph{et~al.}
\newblock \bibinfo{title}{{Spectral Energy Distributions of QSOs at z
  {\ensuremath{>}} 5: Common Active Galactic Nucleus-heated Dust and
  Occasionally Strong Star-formation}}.
\newblock \emph{\bibinfo{journal}{\apj}} \textbf{\bibinfo{volume}{785}},
  \bibinfo{pages}{154} (\bibinfo{year}{2014}).

\bibitem{Wright2010}
\bibinfo{author}{Wright, E.~L.} \emph{et~al.}
\newblock \bibinfo{title}{The wide-field infrared survey explorer (wise):
  mission description and initial on-orbit performance}.
\newblock \emph{\bibinfo{journal}{The Astronomical Journal}}
  \textbf{\bibinfo{volume}{140}}, \bibinfo{pages}{1868} (\bibinfo{year}{2010}).

\bibitem{Tan2024}
\bibinfo{author}{{Tan}, Q.-H.} \emph{et~al.}
\newblock \bibinfo{title}{{Fitting pseudo-S{\'e}rsic (Spergel) light profiles
  to galaxies in interferometric data: The excellence of the
  u{\ensuremath{\upsilon}}-plane}}.
\newblock \emph{\bibinfo{journal}{\aap}} \textbf{\bibinfo{volume}{684}},
  \bibinfo{pages}{A23} (\bibinfo{year}{2024}).

\bibitem{Salvestrini2025}
\bibinfo{author}{{Salvestrini}, F.} \emph{et~al.}
\newblock \bibinfo{title}{{Molecular gas and dust properties in z
  {\ensuremath{>}} 7 quasar hosts}}.
\newblock \emph{\bibinfo{journal}{\aap}} \textbf{\bibinfo{volume}{695}},
  \bibinfo{pages}{A23} (\bibinfo{year}{2025}).

\bibitem{Fujimoto2020}
\bibinfo{author}{{Fujimoto}, S.} \emph{et~al.}
\newblock \bibinfo{title}{{The ALPINE-ALMA [C II] Survey: Size of Individual
  Star-forming Galaxies at z = 4-6 and Their Extended Halo Structure}}.
\newblock \emph{\bibinfo{journal}{\apj}} \textbf{\bibinfo{volume}{900}},
  \bibinfo{pages}{1} (\bibinfo{year}{2020}).

\bibitem{Faisst2020}
\bibinfo{author}{{Faisst}, A.~L.} \emph{et~al.}
\newblock \bibinfo{title}{{The ALPINE-ALMA [C II] Survey: Multiwavelength
  Ancillary Data and Basic Physical Measurements}}.
\newblock \emph{\bibinfo{journal}{\apjs}} \textbf{\bibinfo{volume}{247}},
  \bibinfo{pages}{61} (\bibinfo{year}{2020}).

\bibitem{Ikeda2025}
\bibinfo{author}{{Ikeda}, R.} \emph{et~al.}
\newblock \bibinfo{title}{{The ALMA-CRISTAL Survey: Spatial extent of [CII]
  line emission in star-forming galaxies at z = 4‑6}}.
\newblock \emph{\bibinfo{journal}{\aap}} \textbf{\bibinfo{volume}{693}},
  \bibinfo{pages}{A237} (\bibinfo{year}{2025}).

\bibitem{ZhuangM:2023}
\bibinfo{author}{{Zhuang}, M.-Y.} \& \bibinfo{author}{{Ho}, L.~C.}
\newblock \bibinfo{title}{{Evolutionary paths of active galactic nuclei and
  their host galaxies}}.
\newblock \emph{\bibinfo{journal}{Nature Astronomy}}
  \textbf{\bibinfo{volume}{7}}, \bibinfo{pages}{1376--1389}
  (\bibinfo{year}{2023}).

\bibitem{Xie2024}
\bibinfo{author}{{Xie}, L.} \emph{et~al.}
\newblock \bibinfo{title}{{The First Quenched Galaxies: When and How?}}
\newblock \emph{\bibinfo{journal}{\apjl}} \textbf{\bibinfo{volume}{966}},
  \bibinfo{pages}{L2} (\bibinfo{year}{2024}).

\bibitem{Carnall2023}
\bibinfo{author}{{Carnall}, A.~C.} \emph{et~al.}
\newblock \bibinfo{title}{{A massive quiescent galaxy at redshift 4.658}}.
\newblock \emph{\bibinfo{journal}{\nat}} \textbf{\bibinfo{volume}{619}},
  \bibinfo{pages}{716--719} (\bibinfo{year}{2023}).

\bibitem{Ito2024}
\bibinfo{author}{{Ito}, K.} \emph{et~al.}
\newblock \bibinfo{title}{{Size{\textendash}Stellar Mass Relation and
  Morphology of Quiescent Galaxies at z {\ensuremath{\geq}} 3 in Public JWST
  Fields}}.
\newblock \emph{\bibinfo{journal}{\apj}} \textbf{\bibinfo{volume}{964}},
  \bibinfo{pages}{192} (\bibinfo{year}{2024}).

\bibitem{Graaff2025}
\bibinfo{author}{{de Graaff}, A.} \emph{et~al.}
\newblock \bibinfo{title}{{Efficient formation of a massive quiescent galaxy at
  redshift 4.9}}.
\newblock \emph{\bibinfo{journal}{Nature Astronomy}}
  \textbf{\bibinfo{volume}{9}}, \bibinfo{pages}{280--292}
  (\bibinfo{year}{2025}).

\bibitem{WangT:2024}
\bibinfo{author}{Wang, T.} \emph{et~al.}
\newblock \bibinfo{title}{Black holes regulate cool gas accretion in massive
  galaxies}.
\newblock \emph{\bibinfo{journal}{Nature}}  (\bibinfo{year}{2024}).
\newblock \urlprefix\url{http://dx.doi.org/10.1038/s41586-024-07821-2}.

\bibitem{Onoue2024}
\bibinfo{author}{{Onoue}, M.} \emph{et~al.}
\newblock \bibinfo{title}{{A Post-Starburst Pathway to Forming Massive Galaxies
  and Their Black Holes at z{\ensuremath{>}}6}}.
\newblock \emph{\bibinfo{journal}{arXiv e-prints}}
  \bibinfo{pages}{arXiv:2409.07113} (\bibinfo{year}{2024}).

\bibitem{Di2023}
\bibinfo{author}{{Di}, Y.}, \bibinfo{author}{{Li}, Y.},
  \bibinfo{author}{{Yuan}, F.}, \bibinfo{author}{{Shi}, F.} \&
  \bibinfo{author}{{Caradonna}, M.}
\newblock \bibinfo{title}{{Black hole feeding and feedback in a compact
  galaxy}}.
\newblock \emph{\bibinfo{journal}{\mnras}}  (\bibinfo{year}{2023}).

\bibitem{Hopkins2022}
\bibinfo{author}{{Hopkins}, P.~F.}, \bibinfo{author}{{Wellons}, S.},
  \bibinfo{author}{{Angl{\'e}s-Alc{\'a}zar}, D.},
  \bibinfo{author}{{Faucher-Gigu{\`e}re}, C.-A.} \&
  \bibinfo{author}{{Grudi{\'c}}, M.~Y.}
\newblock \bibinfo{title}{{Why do black holes trace bulges (\& central surface
  densities), instead of galaxies as a whole?}}
\newblock \emph{\bibinfo{journal}{\mnras}} \textbf{\bibinfo{volume}{510}},
  \bibinfo{pages}{630--638} (\bibinfo{year}{2022}).

\bibitem{Dekel2025}
\bibinfo{author}{{Dekel}, A.} \emph{et~al.}
\newblock \bibinfo{title}{{Growth of massive black holes in FFB galaxies at
  cosmic dawn}}.
\newblock \emph{\bibinfo{journal}{\aap}} \textbf{\bibinfo{volume}{695}},
  \bibinfo{pages}{A97} (\bibinfo{year}{2025}).

\bibitem{Baggen2023}
\bibinfo{author}{{Baggen}, J. F.~W.} \emph{et~al.}
\newblock \bibinfo{title}{{Sizes and Mass Profiles of Candidate Massive
  Galaxies Discovered by JWST at 7 {\ensuremath{{\ensuremath{<}}}} z
  {\ensuremath{{\ensuremath{<}}}} 9: Evidence for Very Early Formation of the
  Central 100 pc of Present-day Ellipticals}}.
\newblock \emph{\bibinfo{journal}{\apjl}} \textbf{\bibinfo{volume}{955}},
  \bibinfo{pages}{L12} (\bibinfo{year}{2023}).

\bibitem{Furtak2023}
\bibinfo{author}{{Furtak}, L.~J.} \emph{et~al.}
\newblock \bibinfo{title}{{JWST UNCOVER: Extremely Red and Compact Object at z
  $_{phot}$ {\ensuremath{\approx}} 7.6 Triply Imaged by A2744}}.
\newblock \emph{\bibinfo{journal}{\apj}} \textbf{\bibinfo{volume}{952}},
  \bibinfo{pages}{142} (\bibinfo{year}{2023}).

\bibitem{Chen2025}
\bibinfo{author}{{Chen}, C.-H.}, \bibinfo{author}{{Ho}, L.~C.},
  \bibinfo{author}{{Li}, R.} \& \bibinfo{author}{{Zhuang}, M.-Y.}
\newblock \bibinfo{title}{{The Host Galaxy (If Any) of the Little Red Dots}}.
\newblock \emph{\bibinfo{journal}{\apj}} \textbf{\bibinfo{volume}{983}},
  \bibinfo{pages}{60} (\bibinfo{year}{2025}).

\bibitem{Martin2018}
\bibinfo{author}{{Mart{\'\i}n-Navarro}, I.}, \bibinfo{author}{{Brodie}, J.~P.},
  \bibinfo{author}{{Romanowsky}, A.~J.}, \bibinfo{author}{{Ruiz-Lara}, T.} \&
  \bibinfo{author}{{van de Ven}, G.}
\newblock \bibinfo{title}{{Black-hole-regulated star formation in massive
  galaxies}}.
\newblock \emph{\bibinfo{journal}{\nat}} \textbf{\bibinfo{volume}{553}},
  \bibinfo{pages}{307--309} (\bibinfo{year}{2018}).

\bibitem{Bosch2016}
\bibinfo{author}{{van den Bosch}, R. C.~E.}
\newblock \bibinfo{title}{{Unification of the fundamental plane and Super
  Massive Black Hole Masses}}.
\newblock \emph{\bibinfo{journal}{\apj}} \textbf{\bibinfo{volume}{831}},
  \bibinfo{pages}{134} (\bibinfo{year}{2016}).

\bibitem{Dekel2023}
\bibinfo{author}{{Dekel}, A.}, \bibinfo{author}{{Sarkar}, K.~C.},
  \bibinfo{author}{{Birnboim}, Y.}, \bibinfo{author}{{Mandelker}, N.} \&
  \bibinfo{author}{{Li}, Z.}
\newblock \bibinfo{title}{{Efficient formation of massive galaxies at cosmic
  dawn by feedback-free starbursts}}.
\newblock \emph{\bibinfo{journal}{\mnras}} \textbf{\bibinfo{volume}{523}},
  \bibinfo{pages}{3201--3218} (\bibinfo{year}{2023}).

\bibitem{Grudic2018}
\bibinfo{author}{{Grudi{\'c}}, M.~Y.} \emph{et~al.}
\newblock \bibinfo{title}{{When feedback fails: the scaling and saturation of
  star formation efficiency}}.
\newblock \emph{\bibinfo{journal}{\mnras}} \textbf{\bibinfo{volume}{475}},
  \bibinfo{pages}{3511--3528} (\bibinfo{year}{2018}).

\bibitem{Tanaka2024}
\bibinfo{author}{{Tanaka}, M.} \emph{et~al.}
\newblock \bibinfo{title}{{A Protocluster of Massive Quiescent Galaxies at z =
  4}}.
\newblock \emph{\bibinfo{journal}{\apj}} \textbf{\bibinfo{volume}{970}},
  \bibinfo{pages}{59} (\bibinfo{year}{2024}).

\bibitem{Setton2024}
\bibinfo{author}{{Setton}, D.~J.} \emph{et~al.}
\newblock \bibinfo{title}{{UNCOVER NIRSpec/PRISM Spectroscopy Unveils Evidence
  of Early Core Formation in a Massive, Centrally Dusty Quiescent Galaxy at
  $z_{spec}=3.97$}}.
\newblock \emph{\bibinfo{journal}{arXiv e-prints}}
  \bibinfo{pages}{arXiv:2402.05664} (\bibinfo{year}{2024}).

\bibitem{Kroupa2001}
\bibinfo{author}{{Kroupa}, P.}
\newblock \bibinfo{title}{{On the variation of the initial mass function}}.
\newblock \emph{\bibinfo{journal}{\mnras}} \textbf{\bibinfo{volume}{322}},
  \bibinfo{pages}{231--246} (\bibinfo{year}{2001}).

\bibitem{Murphy2017}
\bibinfo{author}{{Murphy}, E.~J.} \emph{et~al.}
\newblock \bibinfo{title}{{The GOODS-N Jansky VLA 10 GHz Pilot Survey: Sizes of
  Star-forming {\ensuremath{\mu}}JY Radio Sources}}.
\newblock \emph{\bibinfo{journal}{\apj}} \textbf{\bibinfo{volume}{839}},
  \bibinfo{pages}{35} (\bibinfo{year}{2017}).

\bibitem{Willot2013}
\bibinfo{author}{{Willott}, C.~J.}, \bibinfo{author}{{Omont}, A.} \&
  \bibinfo{author}{{Bergeron}, J.}
\newblock \bibinfo{title}{{Redshift 6.4 Host Galaxies of {}10$^{8}$ Solar Mass
  Black Holes: Low Star Formation Rate and Dynamical Mass}}.
\newblock \emph{\bibinfo{journal}{\apj}} \textbf{\bibinfo{volume}{770}},
  \bibinfo{pages}{13} (\bibinfo{year}{2013}).

\bibitem{DeRosa2014}
\bibinfo{author}{{De Rosa}, G.} \emph{et~al.}
\newblock \bibinfo{title}{{Black Hole Mass Estimates and Emission-line
  Properties of a Sample of Redshift z
  {\ensuremath{>}}{\ensuremath{>}} 6.5 Quasars}}.
\newblock \emph{\bibinfo{journal}{\apj}} \textbf{\bibinfo{volume}{790}},
  \bibinfo{pages}{145} (\bibinfo{year}{2014}).

\bibitem{Willot2015}
\bibinfo{author}{{Willott}, C.~J.}, \bibinfo{author}{{Bergeron}, J.} \&
  \bibinfo{author}{{Omont}, A.}
\newblock \bibinfo{title}{{Star Formation Rate and Dynamical Mass of {}10$^{8}$
  Solar Mass Black Hole Host Galaxies At Redshift 6}}.
\newblock \emph{\bibinfo{journal}{\apj}} \textbf{\bibinfo{volume}{801}},
  \bibinfo{pages}{123} (\bibinfo{year}{2015}).

\bibitem{Wu2015}
\bibinfo{author}{{Wu}, X.-B.} \emph{et~al.}
\newblock \bibinfo{title}{{An ultraluminous quasar with a
  twelve-billion-solar-mass black hole at redshift 6.30}}.
\newblock \emph{\bibinfo{journal}{\nat}} \textbf{\bibinfo{volume}{518}},
  \bibinfo{pages}{512--515} (\bibinfo{year}{2015}).

\bibitem{Willot2017}
\bibinfo{author}{{Willott}, C.~J.}, \bibinfo{author}{{Bergeron}, J.} \&
  \bibinfo{author}{{Omont}, A.}
\newblock \bibinfo{title}{{A Wide Dispersion in Star Formation Rate and
  Dynamical Mass of {}10$^{8}$ Solar Mass Black Hole Host Galaxies at Redshift
  6}}.
\newblock \emph{\bibinfo{journal}{\apj}} \textbf{\bibinfo{volume}{850}},
  \bibinfo{pages}{108} (\bibinfo{year}{2017}).

\bibitem{Izumi2018}
\bibinfo{author}{{Izumi}, T.} \emph{et~al.}
\newblock \bibinfo{title}{{Subaru High-z Exploration of Low-Luminosity Quasars
  (SHELLQs). III. Star formation properties of the host galaxies at z
  {\ensuremath{\gtrsim}} 6 studied with ALMA}}.
\newblock \emph{\bibinfo{journal}{\pasj}} \textbf{\bibinfo{volume}{70}},
  \bibinfo{pages}{36} (\bibinfo{year}{2018}).

\bibitem{Izumi2019}
\bibinfo{author}{{Izumi}, T.} \emph{et~al.}
\newblock \bibinfo{title}{{Subaru High-z Exploration of Low-Luminosity Quasars
  (SHELLQs). VIII. A less biased view of the early co-evolution of black holes
  and host galaxies}}.
\newblock \emph{\bibinfo{journal}{\pasj}} \textbf{\bibinfo{volume}{71}},
  \bibinfo{pages}{111} (\bibinfo{year}{2019}).

\bibitem{Reed2019}
\bibinfo{author}{{Reed}, S.~L.} \emph{et~al.}
\newblock \bibinfo{title}{{Three new VHS-DES quasars at 6.7
  {\ensuremath{{\ensuremath{<}}}} z {\ensuremath{{\ensuremath{<}}}} 6.9 and
  emission line properties at z {\ensuremath{>}} 6.5}}.
\newblock \emph{\bibinfo{journal}{\mnras}} \textbf{\bibinfo{volume}{487}},
  \bibinfo{pages}{1874--1885} (\bibinfo{year}{2019}).

\bibitem{Shen2019}
\bibinfo{author}{{Shen}, Y.} \emph{et~al.}
\newblock \bibinfo{title}{{Gemini GNIRS Near-infrared Spectroscopy of 50
  Quasars at z {\ensuremath{\gtrsim}} 5.7}}.
\newblock \emph{\bibinfo{journal}{\apj}} \textbf{\bibinfo{volume}{873}},
  \bibinfo{pages}{35} (\bibinfo{year}{2019}).

\bibitem{Andika2020}
\bibinfo{author}{{Andika}, I.~T.} \emph{et~al.}
\newblock \bibinfo{title}{{Probing the Nature of High-redshift Weak Emission
  Line Quasars: A Young Quasar with a Starburst Host Galaxy}}.
\newblock \emph{\bibinfo{journal}{\apj}} \textbf{\bibinfo{volume}{903}},
  \bibinfo{pages}{34} (\bibinfo{year}{2020}).

\bibitem{Eilers2020}
\bibinfo{author}{{Eilers}, A.-C.} \emph{et~al.}
\newblock \bibinfo{title}{{Detecting and Characterizing Young Quasars. I.
  Systemic Redshifts and Proximity Zone Measurements}}.
\newblock \emph{\bibinfo{journal}{\apj}} \textbf{\bibinfo{volume}{900}},
  \bibinfo{pages}{37} (\bibinfo{year}{2020}).

\bibitem{Izumi2021a}
\bibinfo{author}{{Izumi}, T.} \emph{et~al.}
\newblock \bibinfo{title}{{Subaru High-z Exploration of Low-luminosity Quasars
  (SHELLQs). XII. Extended [C II] Structure (Merger or Outflow) in a z = 6.72
  Red Quasar}}.
\newblock \emph{\bibinfo{journal}{\apj}} \textbf{\bibinfo{volume}{908}},
  \bibinfo{pages}{235} (\bibinfo{year}{2021}).

\bibitem{Neeleman2021}
\bibinfo{author}{{Neeleman}, M.} \emph{et~al.}
\newblock \bibinfo{title}{{The Kinematics of z {\ensuremath{\gtrsim}} 6 Quasar
  Host Galaxies}}.
\newblock \emph{\bibinfo{journal}{\apj}} \textbf{\bibinfo{volume}{911}},
  \bibinfo{pages}{141} (\bibinfo{year}{2021}).

\bibitem{Izumi2021b}
\bibinfo{author}{{Izumi}, T.} \emph{et~al.}
\newblock \bibinfo{title}{{Subaru High-z Exploration of Low-luminosity Quasars
  (SHELLQs). XIII. Large-scale Feedback and Star Formation in a Low-luminosity
  Quasar at z = 7.07 on the Local Black Hole to Host Mass Relation}}.
\newblock \emph{\bibinfo{journal}{\apj}} \textbf{\bibinfo{volume}{914}},
  \bibinfo{pages}{36} (\bibinfo{year}{2021}).

\bibitem{Yang2021}
\bibinfo{author}{{Yang}, J.} \emph{et~al.}
\newblock \bibinfo{title}{{Probing Early Supermassive Black Hole Growth and
  Quasar Evolution with Near-infrared Spectroscopy of 37 Reionization-era
  Quasars at 6.3 {\ensuremath{<}} z {\ensuremath{\leq}} 7.64}}.
\newblock \emph{\bibinfo{journal}{\apj}} \textbf{\bibinfo{volume}{923}},
  \bibinfo{pages}{262} (\bibinfo{year}{2021}).

\bibitem{Venemans2018}
\bibinfo{author}{{Venemans}, B.~P.} \emph{et~al.}
\newblock \bibinfo{title}{{Dust Emission in an Accretion-rate-limited Sample of
  z {\ensuremath{\gtrsim}} 6 Quasars}}.
\newblock \emph{\bibinfo{journal}{\apj}} \textbf{\bibinfo{volume}{866}},
  \bibinfo{pages}{159} (\bibinfo{year}{2018}).

\bibitem{Venemans2019}
\bibinfo{author}{{Venemans}, B.~P.} \emph{et~al.}
\newblock \bibinfo{title}{{400 pc Imaging of a Massive Quasar Host Galaxy at a
  Redshift of 6.6}}.
\newblock \emph{\bibinfo{journal}{\apjl}} \textbf{\bibinfo{volume}{874}},
  \bibinfo{pages}{L30} (\bibinfo{year}{2019}).

\bibitem{Murphy2011}
\bibinfo{author}{{Murphy}, E.~J.} \emph{et~al.}
\newblock \bibinfo{title}{{Calibrating Extinction-free Star Formation Rate
  Diagnostics with 33 GHz Free-free Emission in NGC 6946}}.
\newblock \emph{\bibinfo{journal}{\apj}} \textbf{\bibinfo{volume}{737}},
  \bibinfo{pages}{67} (\bibinfo{year}{2011}).

\bibitem{Venemans2016}
\bibinfo{author}{{Venemans}, B.~P.} \emph{et~al.}
\newblock \bibinfo{title}{{Bright [C II] and Dust Emission in Three z
  {\ensuremath{>}} 6.6 Quasar Host Galaxies Observed by ALMA}}.
\newblock \emph{\bibinfo{journal}{\apj}} \textbf{\bibinfo{volume}{816}},
  \bibinfo{pages}{37} (\bibinfo{year}{2016}).

\bibitem{Decarli2022}
\bibinfo{author}{{Decarli}, R.} \emph{et~al.}
\newblock \bibinfo{title}{{Molecular gas in z {\ensuremath{\sim}} 6 quasar host
  galaxies}}.
\newblock \emph{\bibinfo{journal}{\aap}} \textbf{\bibinfo{volume}{662}},
  \bibinfo{pages}{A60} (\bibinfo{year}{2022}).

\bibitem{Kaasinen2024}
\bibinfo{author}{{Kaasinen}, M.} \emph{et~al.}
\newblock \bibinfo{title}{{The cold molecular gas in z {\ensuremath{\gtrsim}} 6
  quasar host galaxies}}.
\newblock \emph{\bibinfo{journal}{\aap}} \textbf{\bibinfo{volume}{684}},
  \bibinfo{pages}{A33} (\bibinfo{year}{2024}).

\bibitem{Shen2008}
\bibinfo{author}{{Shen}, Y.}, \bibinfo{author}{{Greene}, J.~E.},
  \bibinfo{author}{{Strauss}, M.~A.}, \bibinfo{author}{{Richards}, G.~T.} \&
  \bibinfo{author}{{Schneider}, D.~P.}
\newblock \bibinfo{title}{{Biases in Virial Black Hole Masses: An SDSS
  Perspective}}.
\newblock \emph{\bibinfo{journal}{\apj}} \textbf{\bibinfo{volume}{680}},
  \bibinfo{pages}{169--190} (\bibinfo{year}{2008}).

\bibitem{Tan2024b}
\bibinfo{author}{{Tan}, Q.-H.} \emph{et~al.}
\newblock \bibinfo{title}{{In-Situ Spheroid Formation in Distant
  Submillimeter-Bright Galaxies}}.
\newblock \emph{\bibinfo{journal}{\nat}} \bibinfo{pages}{arXiv:2407.16578}
  (\bibinfo{year}{2024}).

\bibitem{Yue2021}
\bibinfo{author}{{Yue}, M.} \emph{et~al.}
\newblock \bibinfo{title}{{ALMA Observations of the Sub-kpc Structure of the
  Host Galaxy of a z = 6.5 Lensed Quasar: A Rotationally Supported
  Hyper-Starburst System at the Epoch of Reionization}}.
\newblock \emph{\bibinfo{journal}{\apj}} \textbf{\bibinfo{volume}{917}},
  \bibinfo{pages}{99} (\bibinfo{year}{2021}).

\bibitem{Shangguan2018}
\bibinfo{author}{{Shangguan}, J.}, \bibinfo{author}{{Ho}, L.~C.} \&
  \bibinfo{author}{{Xie}, Y.}
\newblock \bibinfo{title}{{On the Gas Content and Efficiency of AGN Feedback in
  Low-redshift Quasars}}.
\newblock \emph{\bibinfo{journal}{\apj}} \textbf{\bibinfo{volume}{854}},
  \bibinfo{pages}{158} (\bibinfo{year}{2018}).

\bibitem{Seth2014}
\bibinfo{author}{{Seth}, A.~C.} \emph{et~al.}
\newblock \bibinfo{title}{{A supermassive black hole in an ultra-compact dwarf
  galaxy}}.
\newblock \emph{\bibinfo{journal}{\nat}} \textbf{\bibinfo{volume}{513}},
  \bibinfo{pages}{398--400} (\bibinfo{year}{2014}).

\bibitem{FerreMateu2015}
\bibinfo{author}{{Ferr{\'e}-Mateu}, A.}, \bibinfo{author}{{Mezcua}, M.},
  \bibinfo{author}{{Trujillo}, I.}, \bibinfo{author}{{Balcells}, M.} \&
  \bibinfo{author}{{van den Bosch}, R. C.~E.}
\newblock \bibinfo{title}{{Massive Relic Galaxies Challenge the Co-evolution of
  Super-massive Black Holes and Their Host Galaxies}}.
\newblock \emph{\bibinfo{journal}{\apj}} \textbf{\bibinfo{volume}{808}},
  \bibinfo{pages}{79} (\bibinfo{year}{2015}).

\end{thebibliography}

\clearpage
\begin{addendum}
\item [Acknowledgements] 
This work was supported by National Natural Science Foundation of China (Project No.12173017 and Key Project No.12141301), National Key R\&D Program of China (grant no. 2023YFA1605600), Scientific Research Innovation Capability Support Project for Young Faculty (Project No. ZYGXQNJSKYCXNLZCXM-P3
),  and the China Manned Space Program with grant no. CMS-CSST-2025-A04.
\item [Author contributions] 
Y.W. reduced the ALMA data, contributed to the main results, and authored an initial draft of the paper under the supervision of T.W.. T.W. initiated the study, lead the interpretation of the main results, and improved the text. D.L. helped with the ALMA data reduction. Q.T. helped with the size fitting using GILDAS. L.C., Z.Z., Y.S., K.X., K.K., W.R., T.I. and Z.L. contributed to the overall interpretation of the results, various aspects of the analysis and improvements of the text.
\item[Author Information] 
The authors declare no competing interests.
\item [Data Availability] 
All ALMA data used in this work are available via ALMA Archive. We summarize the key physical parameters of this quasar sample in Extended Data Table\ref{etb1}.
\item [Code Availability] 
All codes used in this paper are publicly available. 

\end{addendum}

\end{spacing}
\end{document}